\documentclass[twocolumn]{revtex4-1}
\pdfoutput=1

\usepackage{tikz,amsmath, amssymb,bm,color}
\usetikzlibrary{shapes,arrows}
\usetikzlibrary{calc}

\usetikzlibrary{matrix}
\usetikzlibrary{shapes.multipart}

\tikzstyle{basicnode} = [circle, thick, fill=lightgray!30, draw=black, text=black]
\tikzstyle{node} = [basicnode]
\tikzstyle{basicedge} = [draw=black, thin, text=black]
\tikzstyle{edge} = [basicedge, thick]
\tikzstyle{textnode} = [font=\small, text=black, draw=none]

\tikzstyle{highlightededge} = [draw=black, dashed, thick]

\tikzstyle{bluenode} = [circle, thick, fill=blue!30, draw=blue, text=black]

\tikzstyle{blackedge} = [draw=black, thick, text=black]
\tikzstyle{blueedge} = [draw=blue, thick, dashed, text=black]
\tikzstyle{rededge} = [draw=red, thick, dashed, text=black]

\tikzstyle{blackdiredge} = [->, draw=black, thick, text=black]
\tikzstyle{bluediredge} = [->, draw=blue, thick, dashed, text=black]
\tikzstyle{reddiredge} = [->, draw=red, thick, dashed, text=black]

\tikzstyle{blanknode} = [draw=none]

\usetikzlibrary {quotes}

\usepackage{algorithm}
\usepackage{algpseudocode}
\usepackage{amsfonts}
\usepackage{amsmath}
\usepackage{amssymb}
\usepackage{bbm} 
\usepackage{epigraph}  % Added this for schema definition
\usepackage{epstopdf}
\usepackage{float}
\usepackage{graphicx}
\usepackage{lipsum}
\usepackage{mathtools}
\usepackage{multirow}

\usepackage{svg}

\usepackage{adjustbox}
\usepackage{array}
\usepackage[caption=false,font=footnotesize,labelfont=sf,textfont=sf]{subfig}
\usepackage{textcomp}
\usepackage{stfloats}
\usepackage{url}
\usepackage{verbatim}
\ifpdf
  \DeclareGraphicsExtensions{.eps,.pdf,.png,.jpg}
\else
  \DeclareGraphicsExtensions{.eps} 
\fi

\newcommand{\mnistfig}[1]{
\subfloat[Before]{\includegraphics[width=.20\textwidth]{figures/MNIST_results/subplot_#1-0.png}} \subfloat[After]{\includegraphics[width=.20\textwidth]{figures/MNIST_results/subplot_#1-1.png}}}

\newcommand{\revonetext}[1]{\textcolor{black}{#1}}
\newcommand{\ntext}[1]{\textcolor{black}{#1}}
\newcommand{\ie}{\textit{i.e.}}
\newcommand{\eg}{\textit{e.g.}}
\newcommand{\Prob}{\mathbbm{P}}
\newcommand{\Aut}{\text{Aut}}
\newcommand{\AO}{\text{AO}}
\newcommand{\Stab}{\text{Stab}}

\newcommand{\customalg}{\begin{algorithm}[H]} % Use this version for the arXiv submission

\newcommand{\customfig}{\begin{figure}[h]} % Use this version for the arXiv submission

\begin{document}

\title{SCHENO: Measuring Schema vs. Noise in Graphs}

% REVTEX4-1 versions
\author{Justus Isaiah Hibshman}
% \email{jhibshma@nd.edu}
\affiliation{University of Notre Dame}
\author{Adnan Hoq}
% \email{ahoq@nd.edu}
\affiliation{University of Notre Dame}
\author{Tim Weninger}
% \email{tweninger@nd.edu}
\affiliation{University of Notre Dame}

\begin{abstract}
Real-world data is typically a noisy manifestation of a core pattern (\textit{schema}), and the purpose of data mining algorithms is to uncover that pattern, thereby splitting (\textit{i.e.} decomposing) the data into schema and noise. We introduce SCHENO, a principled evaluation metric for the goodness of a schema-noise decomposition of a graph. SCHENO captures how schematic the schema is, how noisy the noise is, and how well the combination of the two represent the original graph data. We visually demonstrate what this metric prioritizes in small graphs, then show that if SCHENO is used as the fitness function for a simple optimization strategy, we can uncover a wide variety of patterns. Finally, we evaluate several well-known graph mining algorithms with this metric; we find that although they produce patterns, those patterns are not always the best representation of the input data. %Along the way, we prove a theorem relating the symmetries of any pair of $n$-node graphs; this theoretical result informs our algorithm's heuristics. We show that our algorithm can extract a wide variety of patterns from different datasets.
\end{abstract}

\maketitle

\setlength\epigraphwidth{.9\linewidth}
\epigraph{\textbf{Schema:}\cite{oed:schema}\newline 1. \textit{A representation of a plan or theory in the form of an outline or model}\newline
2. \textit{A conception of what is common to all members of a class; a general or essential type or form}}

\section{Introduction}
Humans cannot perceive the world in all its raw detail; there is simply too much information for us to do so. Thus, we see the world though the lens of patterns and structures. When looking at a tree, we do not see each leaf individually but rather see leaves. When looking at a leaf, we do not see each cell individually but rather see its shape, color, texture, and veins.

Holistic perception involves noticing both the schema (or form) of an object as well as some of the ways that the object deviates from the pure form. For example, to holistically perceive a human, one must notice both their humanity (the pattern) and some of their idiosyncrasies---short hair, a square jaw, etc. These deviations are themselves perceived through the lens of sub-structure and sub-deviations rather than in terms of atomic units; for example, we see the sub-structure of beard vs. no-beard rather than individual hairs.%, and idiosyncrasies like beard shape, thickness, etc.
Without these structures to help shape our perceptions, carrying out basic life tasks would be overwhelming.

The same limitation applies to computers when processing even moderately-sized datasets. There is simply too much information for computers to process in all possible detail. To obtain useful outputs in a reasonable amount of time, we must program computers to look at data through the lens of some structure. For instance, many statistical models look at data through the structural assumption of linearity. Bayesian causality analysis looks at data through the lens of conditional probability and directed networks that indicate independence assumptions. Convolutional neural networks process images in terms of the composition of small patterns (aka convolutions) like edges or points. More broadly, though we do not fully understand what neural networks consider, we know that the neural architecture itself imposes structural assumptions on the processing of data (\textit{e.g.}, feed-forward vs LSTM architectures), and the network learns to decompose the chaos of raw data into some implicit set of internal features (\textit{i.e.}, patterns).

This limitation creates a conundrum when we want to program a computer to uncover new phenomena: \emph{How can we tell a computer what to look for without already knowing what the patterns and noise it should look for are like?} To date, neural networks seem to be our best shot at getting around this conundrum, because the use of neural networks imposes only mild assumptions about what patterns to look for, and then an objective that we define tells the neural network how to structure itself further. However, we are generally unable to understand the forms that a neural network uncovers. Despite much effort at interpreting trained networks, they largely remain black boxes, and when understanding \emph{is} gained about the neural network's behavior, the understanding usually comes in the form of a schema or pattern that we could have noticed using a simpler model.

Insofar as it is possible, the present work sets out to tackle this conundrum in the context of graphs. We ask: When looking at network data, what is best perceived as the core pattern (schema), and what is best perceived as the noise? The answer to this question will always depend, in part, on the specific task. When the task is to predict new links, the relevant schema is whatever facets of the graph enable that prediction, and the noise is whatever data is not useful. When the task is to detect fraudulent transactions, the structures to find are the fraudulent patterns, and the rest of the data is noise (or alternatively, the structure is that of ordinary transactions and fraud is the noisy exception).

But what if the task is simply that of curiosity? What if we, as scientists exploring the world, simply want to find \textit{the schema} and \textit{the noise} without reference to any particular goal---patterns that might one-day be useful for a goal, but which also are of interest on their own? Is that even possible? In the present work, we argue that the answer is yes, this is possible via the following contributions:

\begin{enumerate}
    \item We introduce a \textit{schema-noise decomposition} task. For any graph, we consider partitioning its edges and non-edges into \textit{schema} and \textit{noise}. That is, every edge or non-edge is considered part of the pattern or part of the noise; we call this partitioning a \textit{schema-noise decomposition} of the graph.
    \item We provide a principled, goal-agnostic definition of pattern and of noise in graphs. Given these definitions, we can \emph{quantify} the goodness of a schema-noise decomposition. In other words, we provide a scoring function for these decompositions. We call our scoring function SCHENO (for SCHEma-NOise). \revonetext{SCHENO uniquely balances three separate considerations: How patterned is the schema, how noisy is the noise, and how well does the proposed schema represent the actual graph?}
    \item We use SCHENO to analyze the performance of four landmark graph mining models: Vocabulary of Graphs (VoG)~\cite{koutra2014vog}, SUBDUE~\cite{cook1993substructure}, the $k$-truss~\cite{cohen2008trusses}, \ntext{and Graph Isomorphism Networks (GINs)~\cite{xu2018powerful}}. Our analysis indicates that although these models can extract real patterns, it is sometimes questionable as to whether those patterns truly represent the original graph. %One can always find some patterned sub-structure, even in a tangled mess of data, but that does not always mean that one has found a legitimate schema for the graph.
    \item Finally, we provide an algorithm to search for good decompositions, thereby enabling us to discover new and interesting schemas in whatever data we happen to care about. %Along the way, we prove a fascinating result concerning how any two graphs' symmetries are related; this result and the logic behind it help speed up our computation and provide a heuristic for which edges in a graph are most likely to be noise.
\end{enumerate}

The new scoring function, SCHENO, is the key contribution of the current work. It gives us the ability to reasonably quantify the goodness of any schema-noise decomposition. This in turn allows us to formally define a goal-agnostic pattern-finding task. Our algorithm for actually finding those decompositions is a simple genetic algorithm with SCHENO as its fitness function; we are certain that more efficient algorithms to optimize SCHENO scores can be developed.

The goal of this paper is, in essence, the quantification of a qualitative concept. Success for us is not about achieving a quantitative score on a downstream task; our goal is more foundational and philosophical. There is no quantitative way to measure our success, because our primary goal is to \emph{define} success for the pattern-finding task. Instead, we define success in a principled, elegant, intuitive, and general fashion, and our definition rewards qualitatively sensible results. We define SCHENO in a very principled manner, and we show that for a variety of synthetic and real-world datasets, SCHENO incentives graph decompositions that, to our eyes, make qualitative sense.

%To that end, we discuss related work in Section~\ref{sec:related_work} and introduce preliminary formalisms in Section~\ref{sec:prelim}. Then, in Section~\ref{sec:schema} we discuss our notions of schema and disorder. We develop these notions into a function for quantifying the goodness of decompositions of a graph into schema and noise in Section~\ref{sec:objective}.
%In Section~\ref{sec:third_party}, we report how SCHENO measures the results of some well-known graph-mining algorithms.
%In Section~\ref{sec:formal_result} we prove a theorem about graph symmetries useful for our computation and beautiful in its own right.
%Then we let SCHENO guide pattern-discovery with a genetic algorithm discussed in Sections~\ref{sec:algorithm} and~\ref{sec:experiments}, wherein we find that SCHENO can prioritize a wide variety of patterns. Lastly, we offer some concluding thoughts in Section~\ref{sec:discussion}.

\section{Related Work}\label{sec:related_work}

The closest attempts that we know of to defining pattern or structure in graphs are the attempts to define entropy for graphs. When Claude Shannon famously defined entropy, he defined it in terms of a probability distribution~\cite{shannon1948mathematical}; entropy measures the overall amount of uncertainty represented by the probability distribution---and conversely the amount of information one gets on average when sampling from the distribution. However, it is very tricky, if not impossible, to translate this idea to graphs. To our knowledge, most attempts to generate an entropy measure for graphs involve first converting a graph to a probability distribution and then using the standard entropy definition for that distribution~\cite{mowshowitz2012entropy, wen2019measuring}; the problem is that in all these cases, some of the information about the graph is lost in the conversion. For example, one might use the node degrees or the sizes of their automorphism orbits to get a distribution~\cite{dehmer2011generalized,dehmer2010inequalities,xiao2008symmetry}, but one cannot reconstruct the graph from this information~\cite{li2016structural}. Li and Pan define entropy over graphs in terms of information needed to communicate where in a graph one is during a random walk, and in particular they do so with a hierarchical decomposition of the graph into communities and sub-communities; they also suggest that such a hierarchical decomposition represents the underlying pattern of the graph~\cite{li2016structural,sikdar2019modeling}. In our own work, we want to avoid making a hierarchical assumption about the layout of the data, even though such an assumption is good for a wide variety of networks.

In a related vein, Choi and Szpankowski consider the entropy of the Erd\H os-R\'enyi distribution over graphs and look into compressing random graphs (up to isomorphism)~\cite{choi2012compression}. Lastly, several works have used notions of graph entropy to attempt link prediction~\cite{parisi2018entropy,yin2017evidential,xu2017entropy}.

As we explore in the sections below, our methodology concludes that schema and pattern in graphs is tied to automorphic symmetry. Thus, works that search for near-symmetries in graphs or define structure relative to symmetry are very relevant. Several such projects exist. % For instance, Xiao et. al. explore the task of finding a network ``skeleton'' of a network without ``redundancies'' due to symmetry~\cite{xiao2008network}.
For instance, Fox, Long, and Porteous explore symmetry-finding through the edge contraction operation~\cite{fox2007discovering, porteous2004identification}. In our work, the relevant operations are edge deletion and edge addition. Knueven, Ostrowski, and Pokutta offer a branch-and-bound algorithm to modify a graph so as to make nodes enter the same automorphism orbits as each other~\cite{knueven2018detecting}. This notion of symmetry is related-to but distinct-from the notion we employ; our notion (automorphisms) is more fine-grained. Markov explores various properties of near-symmetries and some ways to find them~\cite{markovalgebraic,markov2007almost}. Unlike these works, our search goal considers not only how much symmetry is gained, but how much the change required to gain the symmetry looks like random noise. Nevertheless, we suspect that using some of the ideas and heuristics from these works might be beneficial for an algorithm attempting to decompose a graph into schema and noise.% as we define mathematically below.

Some have explored symmetry in real-world networks, finding that such networks have much more symmetry than random graphs~\cite{ball2018symmetric,macarthur2008symmetry}.

\revonetext{SCHENO defines a goal when searching for schemas. We consider other structure-finding goals to be either too specific or too general for the schema-noise decomposition task. Most goals involve searching for particular kinds of sub-structures (\eg, cliques, stars, chains, etc.), each of which bear internal symmetry and thus are encapsulated by the notion of symmetry, but symmetry accounts for other kinds of patterns as well. Other goals involve the repetition of certain small sub-structures, but this does not account for all possible global patterns spanning the graph. If the algorithm-user's goal is to find a specific substructure, then one of these pre-existing goals is the right one, but if one's goal is to find structure in general, then these goals are too narrow.}

\revonetext{The closest objectives to SCHENO (other than graph entropy measures, which we discussed above), are compression-based objectives. The idea behind compression is that if you've found a way to compress the data, then you've found a pattern within the data. Compression is closely tied to repetition, a notion similar to but distinct from symmetry. Symmetry entails repetition, but repetition does not always entail symmetry. Thus compression is in some sense a more general objective than symmetry, but we argue it is both too general and too narrow for the following reasons:}

%\begin{enumerate}
%    \item
(1) \revonetext{For any set of $N$ objects, there is a limit on how many bits one can compress each item into on average: $\log_2(N)$ bits. A method which compresses one element to fewer than $\log_2(N)$ bits will necessarily represent other elements with more than $\log_2(N)$ bits. Adjacency matrices are already a perfectly compressed representation for the space of directed graphs with self-loops; there are $n^2$ bits in an adjacency matrix and $2^{n^2}$ possible graphs. Consequently, any algorithm which successfully compresses ``the right graphs'' or ``the graphs with structure'' or ``graphs seen in real life''  is secretly smuggling in a narrower notion of structure other than compression -- for no method can compress all graphs. If, like Choi and Szpankowski~\cite{choi2012compression}, one decides to only compress graphs up to isomorphism (\ie, ignoring node labels), then one is ``secretly smuggling in'' a notion of structure mathematically equivalent to ours -- except that in Choi and Szpankowski's case, they prioritize compressing the graphs which are the \emph{least} symmetric.}
%    \item

(2) \revonetext{In practice, the amount of compression achieved is measured relative to an original data format. For the reasons given in (1), consistently achieving compression across all graphs relative to that data format means your algorithm has found a weakness in the data format rather than a pattern in your graph. For instance, to succeed on average when compressing graphs originally stored in the adjacency list format, a compression algorithm must find something which is consistently true of graphs but is not true of adjacency lists (or vice versa). While that may tell you something about graphs in general (or adjacency lists), it does not necessarily reveal what is unique about any particular compressed graph.}
% \end{enumerate}

\revonetext{We observed an extreme example of point 2 when analyzing the SUBDUE algorithm~\cite{cook1993substructure}. Occasionally it would manage to ``compress'' the graph by finding a certain repeated substructure: a single edge.}

\begin{figure}[h]
\centering
\adjustbox{max width=\linewidth}{%
\begin{tikzpicture}[scale=0.75]

\node [textnode] at (-4.5,0.15) {\large Example Graph $G$};
\node [textnode] at (4.5, 0.15) {\large Automorphism Orbit of};
\node [textnode] at (4.5,-0.6) {\large Edge Set $\{(1, 5), (5, 6)\}$ in $G$};

\node [textnode] at (0,-5.8) {\large AO$_G\left( \{(1, 5), (5, 6)\} \right) = \{\textcolor{blue}{\textbf{\{(1, 5), (5, 6)\}}},\ \  \textcolor{green!60!black}{\textbf{\{(6, 7), (7, 3)\}}},$};
\node [textnode] at (3, -6.55) {\large $\textcolor{red!80!brown}{\textbf{\{(2, 5), (5, 6)\}}},\ \  \textcolor{gray!80!black}{\textbf{\{(6, 7), (7, 4)\}}}\}$};

\node [textnode] at (0,-7.55) {\large Stab$_G\left( \{(1, 5), (5, 6)\} \right) = \{(1{\rightarrow}1, 2{\rightarrow}2, 3{\rightarrow}3, 4{\rightarrow}4, 5{\rightarrow}5, 6{\rightarrow}6, 7{\rightarrow}7),$};
\node [textnode] at (3.15, -8.3) {\large $(1{\rightarrow}1, 2{\rightarrow}2, 3{\rightarrow}4, 4{\rightarrow}3, 5{\rightarrow}5, 6{\rightarrow}6, 7{\rightarrow}7)\}$};

% Graphs
\begin{scope}[shift={(-4.5,-4)}]
\node [basicnode, label=6] (v6) at (0,1) {};
\node [basicnode, label=7] (v7) at (1.5,1) {};
\node [basicnode, label=5] (v5) at (-1.5,1) {};
\node [basicnode, label=1] (v1) at (-3,2) {};
\node [basicnode, label=2] (v2) at (-3,0) {};
\node [basicnode, label=3] (v3) at (3,2) {};
\node [basicnode, label=4] (v4) at (3,0) {};

\draw [basicedge] (v2) edge (v5);
\draw [basicedge] (v1) edge (v5);
\draw [basicedge] (v5) edge (v6);
\draw [basicedge] (v6) edge (v7);
\draw [basicedge] (v7) edge (v3);
\draw [basicedge] (v7) edge (v4);
\end{scope}

\begin{scope}[shift={(4.5,-4)}]
\node [basicnode, label=\textcolor{lightgray}{6}] (v6) at (0,1) {};
\node [basicnode, label=\textcolor{lightgray}{7}] (v7) at (1.5,1) {};
\node [basicnode, label=\textcolor{lightgray}{5}] (v5) at (-1.5,1) {};
\node [basicnode, label=\textcolor{lightgray}{1}] (v1) at (-3,2) {};
\node [basicnode, label=\textcolor{lightgray}{2}] (v2) at (-3,0) {};
\node [basicnode, label=\textcolor{lightgray}{3}] (v3) at (3,2) {};
\node [basicnode, label=\textcolor{lightgray}{4}] (v4) at (3,0) {};

\draw [basicedge, draw=blue] (v1) edge["a", bend left=15] (v5); % 1-5
\draw [basicedge, draw=blue] (v5) edge["a", bend left=20] (v6); % 5-6

\draw [basicedge, draw=red!80!brown] (v5) edge["b", bend left=15] (v2); % 5-2
\draw [basicedge, draw=red!80!brown] (v6) edge["b", bend left=20] (v5); % 6-5

\draw [basicedge, draw=green!60!black] (v6) edge["c", bend left=20] (v7); % 6-7
\draw [basicedge, draw=green!60!black] (v7) edge["c", bend left=15] (v3); % 7-3

\draw [basicedge, draw=gray!80!black] (v7) edge["d", bend left=20] (v6); % 7-6
\draw [basicedge, draw=gray!80!black] (v4) edge["d", bend left=15] (v7); % 4-7
\end{scope}

% Dividers

\begin{scope}[shift={(9,0)}]
\node [blanknode] (v9) at (-9,-1) {};
\node [blanknode] (v8) at (-9,-5) {};
\draw [basicedge] (v8) edge (v9);
\end{scope}

\end{tikzpicture}
}
\caption{\textbf{Example of the Automorphism Orbit and Stabilizer of a Set of Edges:} On the right, each distinct set of edges in the automorphism orbit is shown in a unique color and labeled with a letter. The multi-edges on the right are shown solely to indicate that the single edges on the left participate in multiple edge sets in the orbit. Note that sets such as $\{(1, 5), (6, 7)\}$ are not part of the example orbit. At the bottom, we can see the two automorphisms in the stabilizer listed; note that the only difference between the two automorphisms is that they swap the positions of nodes $3$ and $4$; everything else is constrained by the stabilized edge set.}\label{fig:example_edge_set_orbit}
\end{figure}

\section{Preliminaries}\label{sec:prelim}

Here we introduce and define several preliminary concepts that are important for our contributions. Many of these definitions and conventions are borrowed from prior work~\cite{hibshman2022ratio}.

\subsection{Notation}\label{sec:shorthand_notation}

Let $S$ be a set and let $f : S \rightarrow S$ be a bijection. Given two elements $a, b \in S$, we use $f((a, b))$ to denote $(f(a), f(b))$.
% Likewise, given a set $S' \subseteq S$ we use $f(S')$ to denote $\{f(x)\ |\ x \in S'\}$. 
Similarly, given a set of pairs $X \subseteq S \times S$ we use $f(X)$ to denote $\{f((c, d))\ |\ (c, d) \in X\}$.

Given two sets $A$ and $B$ we use $A \oplus B$ to denote $(A \setminus B) \cup (B \setminus A)$\footnote{Here $\oplus$ is analogous to the XOR function.}. Similarly, given a graph $G = (V, E)$ and set of node-pairs $P$, we use $G \oplus P$ to denote the graph $(V, E \oplus P)$.
\subsection{Iso- and Auto-morphisms}

Given a graph $G = (V, E)$ and a graph $G' = (V', E')$, an isomorphism between $G$ and $G'$ is a bijection $f : V \rightarrow V'$ such that $(a, b) \in E \leftrightarrow f((a, b)) \in E'$. Graphs $G$ and $G'$ are denoted to be isomorphic by writing $G \cong G'$.

An automorphism of $G$ is simply an isomorphism from $G$ to itself.

Let Aut($G$) denote the set of all automorphisms of $G$.

If a graph $G = (V, E)$ has colored edges, then we require that an automorphism only map edges of the same color to each other. Formally, if we have a set of colors $\mathcal{C}$ and edge coloring function $c: E \rightarrow \mathcal{C}$, then an automorphism of $G$ is a bijection $f: V \rightarrow V$ such that: $e \in E \rightarrow \left( f(e) \in E \land c(e) = c(f(e)) \right)$

\subsection{Automorphism Orbits}

Let $x$ be a graph \textit{entity} where $x$ could be a node ($x \in V$), a node pair ($x \in V \times V$), or a set of node pairs ($x \subseteq V \times V$); $x$ does not need to be an induced subgraph. The automorphism orbit of $x$ is the set of all entities that play the same structural role in $G$ that $x$ does:

$$\text{AO}_G(x) = \{f(x)\ |\ f \in \Aut(G)\}$$

Note that if $x = (a, b)$, $x$ could be an edge in $E$ or $x$ could just as easily be a non-edge. Likewise, if $x$ is a set of node pairs, $x$ could be a set of edges, a set of non-edges, or a mixture. Lastly, note that the automorphism orbit of an entity $x$ contains the same kind of entities that $x$ is, per our notation in Section~\ref{sec:shorthand_notation}; the orbit of a node is a set of nodes, the orbit of an edge is a set of edges, etc. To see an example of the automorphism orbit of a set of edges, see Figure~\ref{fig:example_edge_set_orbit}.

\subsection{Stabilizers}

Graph automorphisms are realignments of a graph with itself; if for some graph entity $x$ in a graph $G$, an automorphism of $G$ realigns that $x$ with itself, then that automorphism is a stabilizer for $x$ in $G$~\cite{rose2009action}. Formally, the stabilizer set of an entity $x$ in a graph $G$ is the set:

$$\text{Stab}_G(x) = \{f\ |\ f \in \Aut(G) \land f(x) = x\}$$

As with automorphism orbits, we can define stabilizers for nodes, edges, sets of edges, etc. To see an example of the stabilizer of a set of edges, see Figure~\ref{fig:example_edge_set_orbit}.

\section{Schema, Chaos, and (A)Symmetry}\label{sec:schema}

In this paper, we tend to use the following words as synonyms: schema, pattern, order, form, structure, and symmetry. This cluster of related concepts is probably best encapsulated by the word schema, which is defined as: ``an outline or model; a conception of what is common to all members of a class; a general or essential type or form''~\cite{oed:schema}. However, given that this word is more obtuse, we will often use one of its synonyms for clarity and for linguistic variety.

By contrast, we use a different set of synonyms to refer to the opposite notion: chaos, randomness, noise, disorder. These are intended to represent the absence or the opposite of the presence of schema, instead being a jumbled mess.

In order to model these concepts in the world of graphs, we define two distributions over graphs in order to quantify the order or disorder of a graph. These distributions are the direct inverses of each other, meaning that if graph A is twice as likely as graph B in the pattern and structure distribution, then graph A is half as likely as graph B in the random chaos distribution. More generally, we define the inverse relationship such that for any pair of graphs, if graph A is $c$ times as likely as graph B in distribution 1, then graph A is $\frac{1}{c}$ times as likely as graph B in distribution 2.

This definition of the inverse is unique, meaning that if one distribution is defined, it automatically entails the other distribution. This can be seen by the fact that the moment you fix the probability of a single element in the inverse distribution, the proportionality constraints immediately entail the probabilities of every other element. Given that the elements must sum to 1, we get the uniqueness of the inverse.

Random chaos tends to be easier to model than structure and pattern due to the ability of making independence assumptions and other such simplifications. To represent random noise or chaos, we choose the Erd\H os-R\'enyi distribution over graphs with probability $\frac{1}{2}$. This distribution corresponds to generating a graph by flipping a 50-50 coin for each possible connection. If the coin comes up heads, the connection is added. If it is tails, the connection is removed.

We choose the Erd\H os-R\'enyi distribution because it is the most intuitive and principled choice for modeling chaos in the realm of graphs. Intuitive, because it is based on independent coin flips. Principled in that the definition is simple and completely chaotic; there is no structural correlation between the presence of any two edges. Furthermore, the Erd\H os-R\'enyi distribution has been widely studied and has been used to prove many combinatorial facts about graphs~\cite{bollobas1998random}.

The opposite of our chaos distribution is our order/structure/form distribution. This distribution, which is the exact inverse of the Erd\H os-R\'enyi chaos distribution, turns out to give graphs a probability proportional to the graph's own internal symmetries (\ie, the number of automorphisms of the graph). Once pondered, this distribution is also an intuitive definition of structure or order. In human perception, pattern and order is closely linked to symmetries. For example, in the visual pattern of a tree, there is a symmetric balance between branches on the left and branches on the right. Similarly, there is symmetry in the fact that any given branch is similar to another one. In most visual patterns, such as a window frame, there are various translations and rotations in 3D space that make the frame symmetric with itself. The connection between symmetry and pattern has been well-documented in psychological research~\cite{lockhead1991perception,palmer1991goodness,attneave1954some,hochberg1953quantitative}, though because some symmetries are especially relevant to 3D space, humans do not see all symmetries equally~\cite{chipman1977complexity,palmer1978orientation,royer1981detection}.

\begin{figure}[h]
\centering
\begin{tikzpicture}[scale=0.9]

\node [textnode] at (-1.2, 1.75) {$\Prob_S(G):$};
\node [textnode] at (-0.3, 1.75) {\large $\frac{3}{45}$};
\node [textnode] at (1.5, 1.75) {\large $\frac{1}{45}$};
\node [textnode] at (3.3, 1.75) {\large $\frac{4}{45}$};
\node [textnode] at (5.1, 1.75) {\large $\frac{2}{45}$};
\node [textnode] at (6.8, 1.75) {\large $\frac{12}{45}$};

    \node [node] at (-0.7, 1.0) (a) {};
    \node [node] at (0.2, 1.0) (b) {};
    \node [node] at (-0.7, 0.2) (c) {};
    \node [node] at (0.2, 0.2) (d) {};
    \node [node] at (1.1, 1.0) (e) {};
    \node [node] at (1.9, 1.0) (f) {};
    \node [node] at (1.1, 0.2) (g) {};
    \node [node] at (1.9, 0.2) (h) {};
    \node [node] at (2.9, 1.0) (i) {};
    \node [node] at (3.7, 1.0) (j) {};
    \node [node] at (2.9, 0.2) (k) {};
    \node [node] at (3.7, 0.2) (l) {};
    \node [node] at (4.7, 1.0) (m) {};
    \node [node] at (5.5, 1.0) (n) {};
    \node [node] at (4.7, 0.2) (o) {};
    \node [node] at (5.5, 0.2) (p) {};
    \node [node] at (6.4, 1.0) (q) {};
    \node [node] at (7.2, 1.0) (r) {};
    \node [node] at (6.4, 0.2) (s) {};
    \node [node] at (7.2, 0.2) (t) {};
    \draw [edge] (a) to (b);
    \draw [edge] (a) to (c);
    \draw [edge] (a) to (d);
    \draw [edge] (e) to (f);
    \draw [edge] (e) to (g);
    \draw [edge] (e) to (h);
    \draw [edge] (f) to (h);
    \draw [edge] (i) to (j);
    \draw [edge] (i) to (k);
    \draw [edge] (j) to (l);
    \draw [edge] (k) to (l);
    \draw [edge] (m) to (n);
    \draw [edge] (m) to (o);
    \draw [edge] (m) to (p);
    \draw [edge] (n) to (p);
    \draw [edge] (o) to (p);
    \draw [edge] (q) to (r);
    \draw [edge] (q) to (s);
    \draw [edge] (q) to (t);
    \draw [edge] (r) to (s);
    \draw [edge] (r) to (t);
    \draw [edge] (s) to (t);
\end{tikzpicture}
\caption{\textbf{Example of Schema Distribution:} The probability that the schema distribution applies to a graph is proportional to the amount of symmetry (structure) in the graph.}\label{fig:schema_dist_example}
\end{figure}
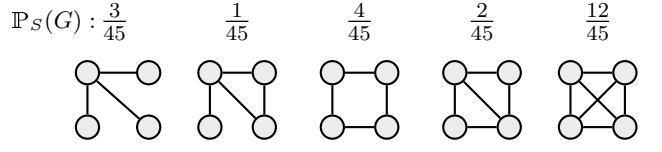

Next, we define these two inverse distributions formally. To get an intuitive sense for the schema distribution, consider the five small graphs depicted in Figure~\ref{fig:schema_dist_example}. The graphs' probabilities are shown next to the graphs. Note that the distribution prioritizes graphs with high degrees of pattern or symmetry over graphs that have little. In addition to making intuitive and theoretical sense, we ultimately find in Section~\ref{sec:experiments} that the structure distribution works well in practice when quantifying what to look for when finding patterns in real-world and synthetic graphs.

Given a graph $G = (V, E)$ with $|V| = n$, $|E| = m$, and the max possible number of edges $M$ (\ie, $\binom{n}{2}$ if $G$ is undirected, $n(n-1)$ if directed), the probability of getting a graph $H \cong G$ with the Erd\H os-R\'enyi distribution parameterized by edge probability $p$ is written as:

\begin{equation}
    \Prob_{\text{ER}_p}(H \cong G) \coloneqq p^m (1 - p)^{M - m} \frac{n!}{|\Aut(G)|}
\end{equation}

For simplicity, we tend to write $\Prob_{\text{ER}_p}(G)$ to denote $\Prob_{\text{ER}_p}(H \cong G)$, treating the randomly sampled graph $H$ as implicit.

When $p = \frac{1}{2}$, as in our \textit{Total Chaos} distribution $\Prob_C$, then we get that:

\begin{equation}
    \Prob_C(H \cong G) \coloneqq \Prob_{\text{ER}_\frac{1}{2}}(G) = \left(\frac{1}{2}\right)^M \cdot \frac{n!}{|\Aut(G)|}
\end{equation}

Let $\mathcal{G}$ be a set containing one of each of the isomorphically distinct graphs on $n$ nodes. Then we get that our schema distribution $\Prob_S$ is defined to be:

\begin{equation}
    \begin{aligned}
        \Prob_S(H \cong G) & \coloneqq \frac{\frac{1}{\Prob_C(H \cong G)}}{\sum_{G' \in \mathcal{G}} \frac{1}{\Prob_C(H \cong G')}} \\
               & = \frac{\frac{1}{\left(\frac{1}{2}\right)^M \cdot \frac{n!}{|\Aut(G)|}}}{\sum_{G' \in \mathcal{G}} \frac{1}{\left(\frac{1}{2}\right)^M \cdot \frac{n!}{|\Aut(G')|}} } \\
               & = \frac{|\Aut(G)|}{\sum_{G' \in \mathcal{G}} |\Aut(G')|}
    \end{aligned}
\end{equation}

Again, we tend to write $\Prob_C(G)$ and $\Prob_S(G)$ rather than $\Prob_C(H \cong G)$ and $\Prob_S(H \cong G)$ respectively, treating the randomly sampled graph $H$ as implicit.

\section{The Formal Objective: SCHENO}\label{sec:objective}

We assume that most real world graphs are noisy realizations of some parsimonious foundational pattern. Our overarching goal is to uncover that pattern. In order to do so, we formalize an objective that quantifies how good a pattern is given the data.

Given a graph $G = (V, E)$, the basic idea is that we assume a \textit{hypothesis graph} $H$ was selected from the schema distribution, and then chaotic noise $N$ was added to $H$ in order to get $G$. Formally, $G = H \oplus N$ (see Section~\ref{sec:shorthand_notation}). Informally, every edge in $N$ that is not in $H$ gets added to $H$, and every edge in $N$ that is also in $H$ is deleted from $H$ -- hence the use of XOR notation $\oplus$.

We search for the (hypothesis-graph and noise) combination that is most likely given the probability distributions for schema and noise.

The probability of $H$ is simply the probability of getting something isomorphic to $H$ when sampling from the schema distribution: $\Prob_S(H)$.

The probability of the noise $N$ is a bit more complex. It is the probability of getting noise isomorphic to $N$ \emph{given} $H$. In other words, it is the probability that $N$ or something equivalent to $N$ in $H$ was the noise (\textit{i.e.}, something in $N$'s automorphism orbit was the noise). The noise is Erd\H os-R\'enyi noise in the sense that any given edge or non-edge is in $N$ independently with some probability $p$. In a slight abuse of notation, we write this as $\Prob_{\text{ER}_p}(N\ |\ H)$, defined to be:

\begin{equation}
    \Prob_{\text{ER}_p}(N\ |\ H) = |\AO_H(N)| \cdot p^{|N|} (1-p)^{M - |N|}
\end{equation}

Where $M$ is the maximum possible number of edges on a graph of $|V|$ nodes. This equation can be thought of as consisting of two parts: 1) the number of noise options structurally equivalent to $N$ in $H$ (\ie\ $|\AO_H(N)|$), and the probability of a single noise set of size $|N|$ (the rest of the equation).

The SCHENO score for an $H$, $N$ pair is then:

\begin{equation}
    \begin{aligned}
    \text{SCH}&\text{ENO}(H, N) \\
    =&\ \Prob_{\text{ER}_p}(N\ |\ H) \cdot \Prob_S(H) \\
    =&\ \frac{|\Aut(H)|}{\sum_{G' \in \mathcal{G}} |\Aut(G')|} \left(|\AO_H(N)| \cdot p^{|N|} (1-p)^{M - |N|} \right)
    \end{aligned}
\end{equation}

As a reminder, $\mathcal{G}$ represents a set containing one of each isomorphically distinct graph on $|V|$ nodes. Since the large summation over $\mathcal{G}$ is the same for all $H$, $N$, we leave it out when actually searching for an optimal schema-noise decomposition. In practice, for computational purposes, we optimize the equivalent objective derived from taking the log and ignoring the values that remain constant:

% \fullp{h}
\begin{figure}[t]
\centering
\adjustbox{max width=\linewidth}{%
\begin{tikzpicture}[scale=1]

% Graph Titles
\node [textnode] at (1, 7.5) {$G$};
\node [textnode] at (6.5, 7.5) {$N_1$};
\node [textnode] at (9.5, 7.5) {$H_1$};
\node [textnode] at (-0.5, 2.5) {$N_2$};
\node [textnode] at ( 2.5, 2.5) {$H_2$};
\node [textnode] at (6.5, 2.5) {$N_3$};
\node [textnode] at (9.5, 2.5) {$H_3$};
\node [textnode] at (-0.5, -2.5) {$N_4$};
\node [textnode] at ( 2.5, -2.5) {$H_4$};
\node [textnode] at (6.5, -2.5) {$N_5$};
\node [textnode] at (9.5, -2.5) {$H_5$};

% Divider
\node [blanknode] at (4.5, 8) (w) {};
\node [blanknode] at (4.5, 2.93) (x) {};
\node [blanknode] at (-1.9, 3.05) (y) {};
\node [blanknode] at (4.62, 3.05) (z) {};
\draw [edge, black] (w) to (x);
\draw [edge, black] (y) to (z);

% Graph Subtext
\node [textnode] at (1, 4.5) {$n = 5; p \approx 0.125737; \binom{n}{2} = 10$};
\node [textnode] at (1, 3.9) {Note: A ``decomposition'' with no};
\node [textnode] at (1, 3.4) {noise gives a score $\propto$ to 0.52};

\node [textnode] at (8, 4.5) {SCHENO Score of $(H_1, N_1)$ is $\propto$ to:};
\node [textnode] at (8, 3.9) {$|\Aut(H_1)|\cdot|\AO_{H_1}(N_1)|\cdot \left(p^1(1-p)^9\right)$};
\node [textnode] at (8, 3.4) {$= 8 \cdot 4 \cdot 0.0375182 \approx 1.2$};

\node [textnode] at (1, -0.5) {SCHENO Score of $(H_2, N_2)$ is $\propto$ to:};
\node [textnode] at (1, -1.1) {$|\Aut(H_2)|\cdot|\AO_{H_2}(N_2)|\cdot \left(p^1(1-p)^9\right)$};
\node [textnode] at (1, -1.6) {$= 12 \cdot 6 \cdot 0.0375182 \approx 2.7$};

\node [textnode] at (8, -0.5) {SCHENO Score of $(H_3, N_3)$ is $\propto$ to:};
\node [textnode] at (8, -1.1) {$|\Aut(H_3)|\cdot|\AO_{H_3}(N_3)|\cdot \left(p^5(1-p)^5\right)$};
\node [textnode] at (8, -1.6) {$= 120 \cdot 60 \cdot 0.000016052 \approx 0.12$};
 
\node [textnode] at (1, -5.5) {SCHENO Score of $(H_4, N_4)$ is $\propto$ to:};
\node [textnode] at (1, -6.1) {$|\Aut(H_4)|\cdot|\AO_{H_4}(N_4)|\cdot \left(p^1(1-p)^9\right)$};
\node [textnode] at (1, -6.6) {$= 2 \cdot 2 \cdot 0.0375182 \approx 0.15$};

\node [textnode] at (8, -5.5) {SCHENO Score of $(H_5, N_5)$ is $\propto$ to:};
\node [textnode] at (8, -6.1) {$|\Aut(H_5)|\cdot|\AO_{H_5}(N_5)|\cdot \left(p^1(1-p)^9\right)$};
\node [textnode] at (8, -6.6) {$= 2 \cdot 1 \cdot 0.0375182 \approx 0.075$};

    \node [node] at (1.0, 7.0) (a) {};
    \node [node] at (0.0, 6.0) (b) {};
    \node [node] at (1.0, 6.0) (c) {};
    \node [node] at (2.0, 6.0) (d) {};
    \node [node] at (1.0, 5.0) (e) {};
    
    \node [node] at (6.5, 7.0) (f1) {};
    \node [node] at (5.5, 6.0) (g1) {};
    \node [node] at (6.5, 6.0) (h1) {};
    \node [node] at (7.5, 6.0) (i1) {};
    \node [node] at (6.5, 5.0) (j1) {};
    \node [node] at (9.5, 7.0) (f2) {};
    \node [node] at (8.5, 6.0) (g2) {};
    \node [node] at (9.5, 6.0) (h2) {};
    \node [node] at (10.5, 6.0) (i2) {};
    \node [node] at (9.5, 5.0) (j2) {};
    
    \node [node] at (-0.5, 2.0) (k1) {};
    \node [node] at (-1.5, 1.0) (l1) {};
    \node [node] at (-0.5, 1.0) (m1) {};
    \node [node] at ( 0.5, 1.0) (n1) {};
    \node [node] at (-0.5, 0.0) (o1) {};
    \node [node] at (2.5, 2.0) (k2) {};
    \node [node] at (1.5, 1.0) (l2) {};
    \node [node] at (2.5, 1.0) (m2) {};
    \node [node] at (3.5, 1.0) (n2) {};
    \node [node] at (2.5, 0.0) (o2) {};
    
    \node [node] at (6.5, 2.0) (p1) {};
    \node [node] at (5.5, 1.0) (q1) {};
    \node [node] at (6.5, 1.0) (r1) {};
    \node [node] at (7.5, 1.0) (s1) {};
    \node [node] at (6.5, 0.0) (t1) {};
    \node [node] at (9.5, 2.0) (p2) {};
    \node [node] at (8.5, 1.0) (q2) {};
    \node [node] at (9.5, 1.0) (r2) {};
    \node [node] at (10.5, 1.0) (s2) {};
    \node [node] at (9.5, 0.0) (t2) {};
    
    \node [node] at (-0.5, -3.0) (u1) {};
    \node [node] at (-1.5, -4.0) (v1) {};
    \node [node] at (-0.5, -4.0) (w1) {};
    \node [node] at (0.5, -4.0) (x1) {};
    \node [node] at (-0.5, -5.0) (y1) {};
    \node [node] at (2.5, -3.0) (u2) {};
    \node [node] at (1.5, -4.0) (v2) {};
    \node [node] at (2.5, -4.0) (w2) {};
    \node [node] at (3.5, -4.0) (x2) {};
    \node [node] at (2.5, -5.0) (y2) {};
    
    \node [node] at (6.5, -3.0) (11) {};
    \node [node] at (5.5, -4.0) (21) {};
    \node [node] at (6.5, -4.0) (31) {};
    \node [node] at (7.5, -4.0) (41) {};
    \node [node] at (6.5, -5.0) (51) {};
    \node [node] at (9.5, -3.0) (12) {};
    \node [node] at (8.5, -4.0) (22) {};
    \node [node] at (9.5, -4.0) (32) {};
    \node [node] at (10.5, -4.0) (42) {};
    \node [node] at (9.5, -5.0) (52) {};

    \draw [edge, black] (a) to (b);
    \draw [edge, black] (a) to (d);
    \draw [edge, black] (b) to (c);
    \draw [edge, black] (b) to (e);
    \draw [edge, black] (d) to (e);
    
    \draw [edge, black] (f2) to (g2);
    \draw [edge, black] (f2) to (i2);
    \draw [edge, red, dashed] (g1) to (h1);
    \draw [edge, black] (g2) to (j2);
    \draw [edge, black] (i2) to (j2);
    
    \draw [edge, black] (k2) to (l2);
    \draw [edge, black] (k2) to (n2);
    \draw [edge, black] (l2) to (m2);
    \draw [edge, black] (l2) to (o2);
    \draw [edge, blue, dashed] (m1) to (n1);
    \draw [edge, black]        (m2) to (n2);
    \draw [edge, black] (n2) to (o2);
    
    \draw [edge, red, dashed] (p1) to (q1);
    \draw [edge, red, dashed] (p1) to (s1);
    \draw [edge, red, dashed] (q1) to (r1);
    \draw [edge, red, dashed] (q1) to (t1);
    \draw [edge, red, dashed] (s1) to (t1);
    
    \draw [edge, black] (u2) to (v2);
    \draw [edge, blue, dashed] (u1) to (w1);
    \draw [edge, black]        (u2) to (w2);
    \draw [edge, black] (u2) to (x2);
    \draw [edge, black] (v2) to (w2);
    \draw [edge, black] (v2) to (y2);
    \draw [edge, black] (x2) to (y2);
    
    \draw [edge, black] (12) to (22);
    \draw [edge, black] (12) to (42);
    \draw [edge, blue, dashed, bend left=30] (11) to (51);
    \draw [edge, black, bend left=30]        (12) to (52);
    \draw [edge, black] (22) to (32);
    \draw [edge, black] (22) to (52);
    \draw [edge, black] (42) to (52);
\end{tikzpicture}
}
\caption{\textbf{Scoring (Schema, Noise) Decompositions:} Given a graph $G$, this figure shows five decompositions of $G$ into Hypothesis Graph (\ie\  Schema) and Noise. The value $p$ is the noise probability -- \ie\ the probability that any edge or non-edge in $G$ used to be a non-edge/edge respectively. To learn how $p$ is calculated as a function of $n$, see Section~\ref{sec:objective}. Black edges are present in both $G$ and $H_i$ but not in $N_i$. Blue dashed edges are the added edges -- present in $H_i$ and $N_i$ but not in $G$. Red dashed edges are the deleted edges -- present in $G$ and $N_i$ but not $H_i$. In all cases, $G = H_i \oplus N_i$.\newline
The values below the graphs are proportional to the scores. The second option gets the highest score; with just one added edge, it manages to greatly increase the amount of symmetry \emph{and} make the noise equivalent to all other edges in the graph. The third option has the most symmetry and the largest set of equivalent noise arrangements, but it incurs a heavy cost for making so many edits and thus gets a very low score. Finally, note that in the last two candidates, the symmetry gain in the graphs is the same (2 automorphisms) but in one case the noise is more probable.}\label{fig:scoring_example}
\end{figure}
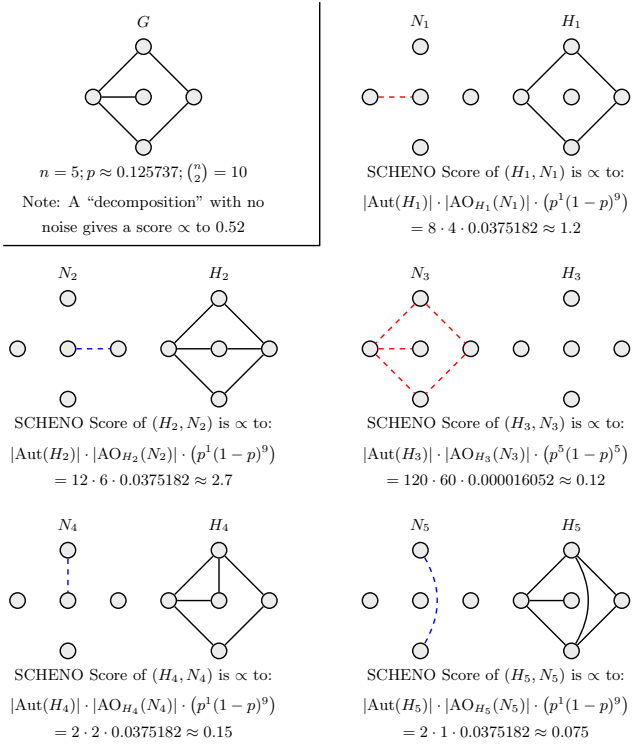

\begin{equation}\label{eqn:score_practice}
\begin{aligned}
    \text{SCH}&\text{ENO Score in Practice}(H, N) \\
    =& \log_2(|\Aut(H)|) + \log_2(|\AO_H(N)|) \\
    +& |N|\log_2(p) + (M - |N|)\log_2(1 - p)
\end{aligned}
\end{equation}

Armed with this equation, we are almost ready to begin searching for structure and noise. There remains one challenge: choosing a value for $p$.

\subsection{Choosing p}\label{sec:choosing_p}

We need to choose some value for our noise probability $p$. We would prefer to do so in a principled manner. We always want $p < \frac{1}{2}$, because $p \geq \frac{1}{2}$ would assume a-priori that each edge is just-as or more likely to have been noise than not, and thus that the graph complement of $G$ would have been just as good or even a better representation of the original structure than $G$.

Given that we will set $p < \frac{1}{2}$, $p$ basically forms a \textit{cost} for making the schema graph different from the data (\ie, a cost on noise). Remember that $\Prob_{\text{ER}_p}(N\ |\ H) = |\AO_H(N)| p^{|N|} (1 - p)^{M-|N|}$. Thus, every element added to the noise set $N$ means one more $p$ being multiplied and one less $(1 - p)$ being multiplied, and because $p < \frac{1}{2} < (1 - p)$, this makes larger noise sets less probable. A value of $p > \frac{1}{2}$ means a score \emph{increase} for more noise, even without symmetry gains. A value of $p = \frac{1}{2}$ means no cost for noise---all that matters is symmetry, and the optimization process is free to wander indefinitely far away from the original graph.

To select $p$ in a principled fashion, let $G_\emptyset = (V, \emptyset)$. We consider two extreme hypotheses: (1) $G$ is the original structure and thus there is no noise, meaning $H = G$ and $N = \emptyset$. (2) The original structure $H$ is the empty graph $G_\emptyset$ and $G$ is all noise, meaning $N = E$. Because one hypothesis is that $G$ is all structure and the other is that $G$ is all noise, we choose to make the relative probability of (1) vs. (2) the same as the extent to which the structure distribution $\Prob_S$ explains $G$ versus the extent to which the chaos distribution $\Prob_{\text{ER}_p}$ explains $G$. This means choosing the value for $p$ that satisfies the following equation:

\begin{equation}
    \frac{\Prob_S(G)}{\Prob_{\text{ER}_p}(G)} = \frac{\text{SCHENO}(G, \emptyset)}{\text{SCHENO}(G_\emptyset, E)}
\end{equation}

Expanding the equation gives us:

\begin{equation}
\begin{aligned}
    & \frac{\frac{|\Aut(G)|}{\sum_{G' \in \mathcal{G}} |\Aut(G')|}}{\frac{n!}{|\Aut(G)|} p^m (1-p)^{M-m}} \\ &= \frac{\frac{|\Aut(G)|}{\sum_{G' \in \mathcal{G}} |\Aut(G')|} |\AO_G(\emptyset)| \cdot p^0 (1-p)^M}  {\frac{|\Aut(G_\emptyset)|}{\sum_{G' \in \mathcal{G}} |\Aut(G')|} |\AO_{G_\emptyset}(E)| \cdot p^m (1-p)^{M-m}}
\end{aligned}
\end{equation}

Given that $|\Aut(G_\emptyset)| = n!$, $|\AO_G(\emptyset)| = 1$ and $|\AO_{G_\emptyset}(E)| = \frac{n!}{|\Aut(G)|}$, simplifying and re-arranging gives us:

\begin{equation}
\begin{aligned}
    & \frac{|\Aut(G)|^2}{n! \cdot p^m \cdot (1-p)^{M-m} \cdot \sum_{G' \in \mathcal{G}} |\Aut(G')|} \\ 
    &= \frac{|\Aut(G)| \cdot p^0 (1-p)^M}{n! \frac{n!}{|\Aut(G)|} \cdot p^m (1-p)^{M-m}}
\end{aligned}
\end{equation}

Further simplification yields:

\begin{equation}
    \frac{n!}{\sum_{G' \in \mathcal{G}} |\Aut(G')|} = (1-p)^M
\end{equation}

Which in turn means that:

\begin{equation}\label{eqn:final_p_formula}
    \log_2(1-p) = \frac{\log_2(n!) - \log_2\left(\sum_{G' \in \mathcal{G}} |\Aut(G')|\right)}{M}
\end{equation}

So far we managed to avoid needing to calculate $\sum_{G' \in \mathcal{G}} |\Aut(G')|$, but now it is important to know. Unfortunately, this value is not known---worse, even $|\mathcal{G}|$ is not known when $n$ is larger than 20 or so. What we do know is that $\sum_{G' \in \mathcal{G}} |\Aut(G')| > |\mathcal{G}| > \frac{2^M}{n!}$.

Fortunately, we can calculate the value for very small $n$%($n \leq 8$ for undirected graphs, $n \leq 6$ for directe graphs)
, and there is a reasonable way to estimate the value for larger $n$.

Assume for simplification that all graphs on $n$ nodes have the same automorphism group size $A_n$. Then we get that $\sum_{G' \in \mathcal{G}} |\Aut(G')| = |\mathcal{G}| \cdot A_n$. We also know from combinatorics that $\sum_{G' \in \mathcal{G}} \frac{n!}{|\Aut(G')|} = 2^M$. If $|\Aut(G')|$ always equals $A_n$, then this equation becomes $\frac{|\mathcal{G}| \cdot n!}{2^M} = A_n$. Thus, if we can calculate $|\mathcal{G}|$, then we can calculate $A_n$ and use that to estimate $\sum_{G' \in \mathcal{G}} |\Aut(G')|$ to be $\frac{(|\mathcal{G}|)^2}{\frac{2^M}{n!}}$.

Fortunately, we have fairly tight Big-O bounds for $|\mathcal{G}|$~\cite{harary2014graphical}. For undirected graphs, the number of graphs on $n$ nodes has the big-O upper-bound of:

\begin{equation}
\begin{aligned}
    |\mathcal{G}| & (\text{undirected}) = \frac{2^{\binom{n}{2}}}{n!} \cdot \left[1 + \frac{n(n-1)}{2^{n-1}} \right. \\ &+  \left. \frac{(n!)}{(n-4)!}\frac{(3n-7)/(3n-9)}{2^{2n-3}} + O\left(\frac{n^5}{2^{5n/2}}\right) \right]
\end{aligned}
\end{equation}

For directed graphs, the bound is somewhat similar:

\begin{equation}
\begin{aligned}
    |\mathcal{G}| & (\text{directed}) = \frac{2^{n^2 - n}}{n!} \cdot \left[1 + \frac{n(n-1)}{2^{2n-2}} \right. \\ &+ \left.\frac{(n!)}{(n-4)!}\frac{(3n-7)/(3n-9)}{2^{4n-7}} + O\left(\frac{n^5}{2^{5n}}\right) \right]
\end{aligned}
\end{equation}

As $n$ gets larger, these values get closer and closer to $\frac{2^M}{n!}$. In other words, the fraction of graphs that are rigid (\ie, $|\Aut(G)| = 1$) approaches 100\%~\cite{harary2014graphical}, which means that the margin of error on our estimate will be small.

\subsection{Summarizing SCHENO's Balancing Act}

To fulfill its aims, SCHENO needs to balance three things when it scores a schema-noise decomposition:

\begin{enumerate}
    \item How structured is the schema?
    \item How random is the noise?
    \item How different is the schema from the original graph (i.e. how \emph{much} noise is present)?
\end{enumerate}

SCHENO manages to balance these factors by imagining a two-stage probabilistic process for generating a graph $G$. In stage 1, a schema graph $H$ is sampled from the schema distribution---a distribution which is the exact probabilistic inverse of the total-chaos Erd\H os-R\'enyi distribution. The schema distribution prioritizes automorphic symmetry. In stage 2, some amount of noise $N$ modifies the schema. In particular, every edge becomes a non-edge with probability $p$ and every non-edge becomes an edge with probability $p$.

SCHENO considers everything up to isomorphism, meaning that in stage 1 it considers the probability of generating a schema which is isomorphic to $H$. Similarly, for stage 2, it uses the probability of adding noise which is isomorphic to $N$; this isomorphic equivalence of noise is defined relative to the schema $H$ under consideration.

The noise probability $p$ is chosen to perfectly balance two extreme hypotheses: the hypothesis that the graph is solely comprised of schema and the hypothesis that the graph is solely comprised of noise. We want the ratio of the SCHENO scores for the all-structure hypothesis vs. the all-noise hypothesis to equal the ratio of the probability of drawing the original graph from the schema distribution vs. the probability of drawing it from an Erd\H os-R\'enyi distribution parametrized by the very same $p$. Ultimately this means that $p$ turns out to be a function of the number of nodes in the graph (and whether or not the graph is directed); the same $p$ acheives this balance for \emph{all} (un)directed graphs on the same number of nodes.

To see SCHENO's balancing in action, consider Fig.~\ref{fig:scoring_example}, which shows five different schema-noise decompositions and the relevant factors determining their SCHENO scores.

% \section{Symmetry Ratio Relating Two $n$-Node Graphs}\label{sec:formal_result}

% \input{math_proof_section}

\section{Analyzing Other Graph Models}\label{sec:third_party}

We evaluate several landmark pattern-finding algorithms for graphs: the $k$-truss~\cite{cohen2008trusses}, SUBDUE~\cite{cook1993substructure}, and Vocabulary of Graphs (VoG)~\cite{koutra2014vog}, \ntext{as well as Graph Isomorphism Networks (GINs)~\cite{xu2018powerful}.} These widely-used algorithms represent different paradigms of pattern-finding in graphs.

\begin{table}[t]
    \centering
    \caption{\textbf{Graph Datasets}}
    \label{tab:graphs}
    \begin{tabular}{llcrr}
    \hline
    \textbf{Graph} & \textbf{Type} & (\textbf{U}n)\textbf{D}irected &  \textbf{Nodes} & \textbf{Edges} \\ \hline
    Karate          & Social                & U & 34      & 78 \\
    Football        & Competition           & U & 195     & 804 \\
    Foodweb         & Ecological            & D & 183     & 2,476 \\
    Political Blogs & Social                & D & 1,224   & 19,022 \\
    EU-Core         & Email                 & D & 1,005   & 24,929 \\
    Cora            & Citations             & D & 2,708   & 5,429  \\ \hline
    Enron           & Email                 & D & 36,692  & 183,831 \\
    Flickr          & Image Sim             & U & 105,938 & 2,316,948 \\
    Epinions        & Trust                 & D & 75,879  & 405,740 \\ \hline
    Fullerene c180  & Molecular             & U & 180     & 270 \\
    Fullerene c720  & Molecular             & U & 720     & 1,080 \\
    Fullerene c6000 & Molecular             & U & 6,000   & 9,000 \\ \hline
    \end{tabular}
\end{table}

$K$-trusses are meant to represent the core of a graph. Formally, the $k$-truss of a graph is the largest subset of edges within the graph such that every edge in the subset is part of at least $(k - 2)$ triangles. SUBDUE searches for the subgraphs that allow it to compress the original graph. The idea is that the subgraphs which best represent the graph's patterns should allow for the most compression. VoG searches a graph for certain pre-defined types of sub-structures, those in its \textit{vocabulary}: stars, cliques, chains, bipartite cores, near-cliques, and near-bipartite cores. VoG only runs on undirected graphs, so all directed graphs are treated as undirected in experiments. \ntext{Graph Isomorphism Networks (GINs) are a kind of graph neural network (GNN); like other GNNs, GINs aim to learn an implicit representation of the graph in a compressed, abstract feature space which can then be decoded back into the graph or used for other tasks. Unlike the other models, GINs do not directly produce a discrete set of edges which could be used as a schema. Instead they produce a ranking over edges. Thus, rather than trying to produce a single schema-noise decomposition from the GIN's output, we consider the whole range of possible schemas produced by taking the top $k$ ranked edges; we vary $k$ from 0 (empty schema) to $2|E|$ -- twice as many edges as the original graph. For information on our parametrization of the GINs, see Section~\ref{sec:GIN_parameters} in the appendix.}

We run every algorithm on the 6 different real-world graphs described in Table~\ref{tab:graphs}. Then we also run VoG on Enron, Flickr, and Epinions (graphs from the original VoG paper). Similarly, we run SUBDUE on the Fullerene graphs because SUBDUE was originally tested on molecular graphs.

Recall that a SCHENO score is conceptually a probability: How likely is this schema to have been the underlying structure and how likely is this noise to have been added-to/deleted-from that structure? Thus, given two decompositions $A$ and $B$, we can think of them as hypotheses concerning what the underlying pattern of our graph is. The ratio of $A$'s SCHENO score to $B$'s SCHENO score tells us how much more (or less) likely $A$ is than $B$ as a reasonable hypothesis.

We compare the SCHENO score of the graph mining algorithms' decompositions to the score we would have obtained if the graph was left untouched (\ie, the score of the decomposition: schema = $G$, noise = $\emptyset$). A positive ratio therefore indicates whether the graph mining algorithm found a good underlying pattern to represent the graph.

Most of the real-world graphs are sparse and/or asymmetric enough that SCHENO considers the graph to be more noise than structure. Thus randomly deleting a large number of edges from the graph will typically result in a SCHENO score improvement. A good schema-noise decomposition should do better than this haphazard decomposition selection. To make sure that any improvements the algorithms show are due to more than merely deleting some edges, we also compare how more (or less) likely the algorithm's decomposition is than a random decomposition with the same noise size.

\begin{table}[h]
    \centering
    \footnotesize
    \caption{\ntext{\textbf{Likelihood of Graph models to trivial and random decompositions.}}}
    \label{tab:models}
    \begin{tabular}{lllll}
    \hline
     & \textbf{Dataset} & \textbf{\# Decomp.} & \textbf{Gain over} & \textbf{Gain over} \\
\multirow{9}{*}{\rotatebox{90}{$K$-Truss}} &     &  & \textbf{`All Structure'} & \textbf{Random} \\ \hline
& Karate   &  $2^{561}$       &  $2^{-1.2}$   &  $2^{11.4}$ \\
& Football &  $2^{18,915}$    &  $2^{654}$    &  $2^{624}$ \\
& Foodweb  &  $2^{33,306}$    &  $2^{39}$     &  $2^{219}$ \\
& PolBlogs &  $2^{1,496,952}$ &  $2^{2,660}$  &  $2^{2,665}$ \\
& EUCore   &  $2^{1,009,020}$ &  $2^{1,295}$  &  $2^{1,295}$ \\
& Cora     &  $2^{7,330,556}$ &  $2^{21,470}$ &  $2^{14,817}$ \\ \hline 
\multirow{9}{*}{\rotatebox{90}{SUBDUE}} & Karate   &  $2^{561}$       &  $2^{-31}$    &  $2^{-1}$ \\
& Football &  $2^{18,915}$    &  $2^{376}$    &  $2^{-51}$ \\
& Foodweb  &  $2^{33,306}$    &  $2^{84}$     &  $2^{-25}$ \\
& PolBlogs &  $2^{1,496,952}$ &  $2^{8,290}$  &  $2^{-2,474}$ \\
& EUCore   &  $2^{1,009,020}$ &  $2^{4,462}$  &  $2^{-2,498}$ \\
& Cora     &  $2^{7,330,556}$ &  $2^{10,370}$ &  $2^{-3,689}$ \\ 
& Fullerene c180  &  $2^{16,110}$     &  $2^{-9}$    &  $2^{7}$ \\
& Fullerene c720  &  $2^{258,840}$    &  $2^{2}$     &  $2^{-44}$ \\
& Fullerene c6000 &  $2^{17,997,000}$ &  $2^{1,262}$ &  $2^{-1,262}$ \\ \hline
\multirow{9}{*}{\rotatebox{90}{VoG}} & Karate   & $2^{561}$       &  ---          &  --- \\
& Football & $2^{18,915}$    &  $2^{1,602}$  &  $2^{1,044}$ \\
& Foodweb  & $2^{16,653}$    &  $2^{-101}$   &  $2^{401}$ \\
& PolBlogs & $2^{748,476}$   &  $2^{1,558}$  &  $2^{1,249}$ \\
& EUCore   & $2^{504,510}$   &  $2^{1,388}$  &  $2^{1,146}$ \\
& Cora     & $2^{3,665,278}$ &  $2^{38,110}$ &  $2^{11,770}$ \\ 
& Flickr   & $2^{5,611,376,953}$ &  $2^{1,509,513}$ &  $2^{1,529,636}$ \\
& Epinions & $2^{2,878,773,381}$ &  $2^{788,882}$   &  $2^{598,352}$ \\
& Enron    & $2^{673,133,086}$   &  $2^{159,378}$   &  $2^{108,971}$ \\
    \end{tabular}
    \vspace{0cm}
\end{table}

\ntext{The performance of the $K$-truss, SUBDUE, and VoG algorithms in graph decomposition was evaluated using SCHENO in Table~\ref{tab:models}, focusing on two metrics: Gain over All Structure and Gain over Random. The former measures how much more likely SCHENO finds an algorithm’s decomposition compared to treating the entire graph as structure with no noise (Schema = Graph, Noise = $\emptyset$). The latter assesses how much more likely the decomposition is compared to a random decomposition with equivalent noise. These metrics help determine whether the algorithms generate meaningful decompositions beyond simple edge deletion, a baseline that can sometimes yield higher scores when SCHENO considers the graph predominantly noise. The total number of possible decompositions contextualizes the reported likelihoods.}

\ntext{$K$-truss consistently outperforms both trivial and random decompositions, demonstrating its ability to identify meaningful structures. The VoG algorithm also performs well overall but struggles with specific graph types, such as food webs. In food web graphs, where nodes represent species and edges represent predator-prey relationships, VoG fails to capture key patterns. These graphs often consist of overlapping bipartite cores, which VoG seems unable to detect, despite these structures being within its vocabulary. For example, VoG retains 54\% of the food web graph’s edges as structure and discards the remaining 46\%, indicating its limitations on such data.}

\ntext{SUBDUE, by contrast, fares poorly. Its decompositions typically lose to random decompositions of similar size and often consist of disconnected tiny subgraphs that lack meaningful internal structure. While SUBDUE identifies repeated small substructures, it fails to account for how these substructures connect to form the overall graph. This remains true even when applied to highly structured graphs, such as Fullerene molecules. SCHENO appears to value decompositions that emphasize dense cores and fringe nodes, such as those identified by the $K$-truss, over SUBDUE’s approach of isolating repeated subgraphs.}

\ntext{This distinction highlights the importance of considering global patterns. SUBDUE's focus on frequent subgraphs provides insight into local structures but insufficiently captures the overall graph's connectivity. In contrast, VoG searches for larger, highly patterned substructures, aligning more closely with SCHENO's criteria for meaningful decomposition. Thus, VoG stands as an intermediate approach, bridging the gap between the overly simplistic $K$-truss and the overly localized focus of SUBDUE.}

\begin{figure}[t]
\centering
\includegraphics[width=0.9\linewidth]{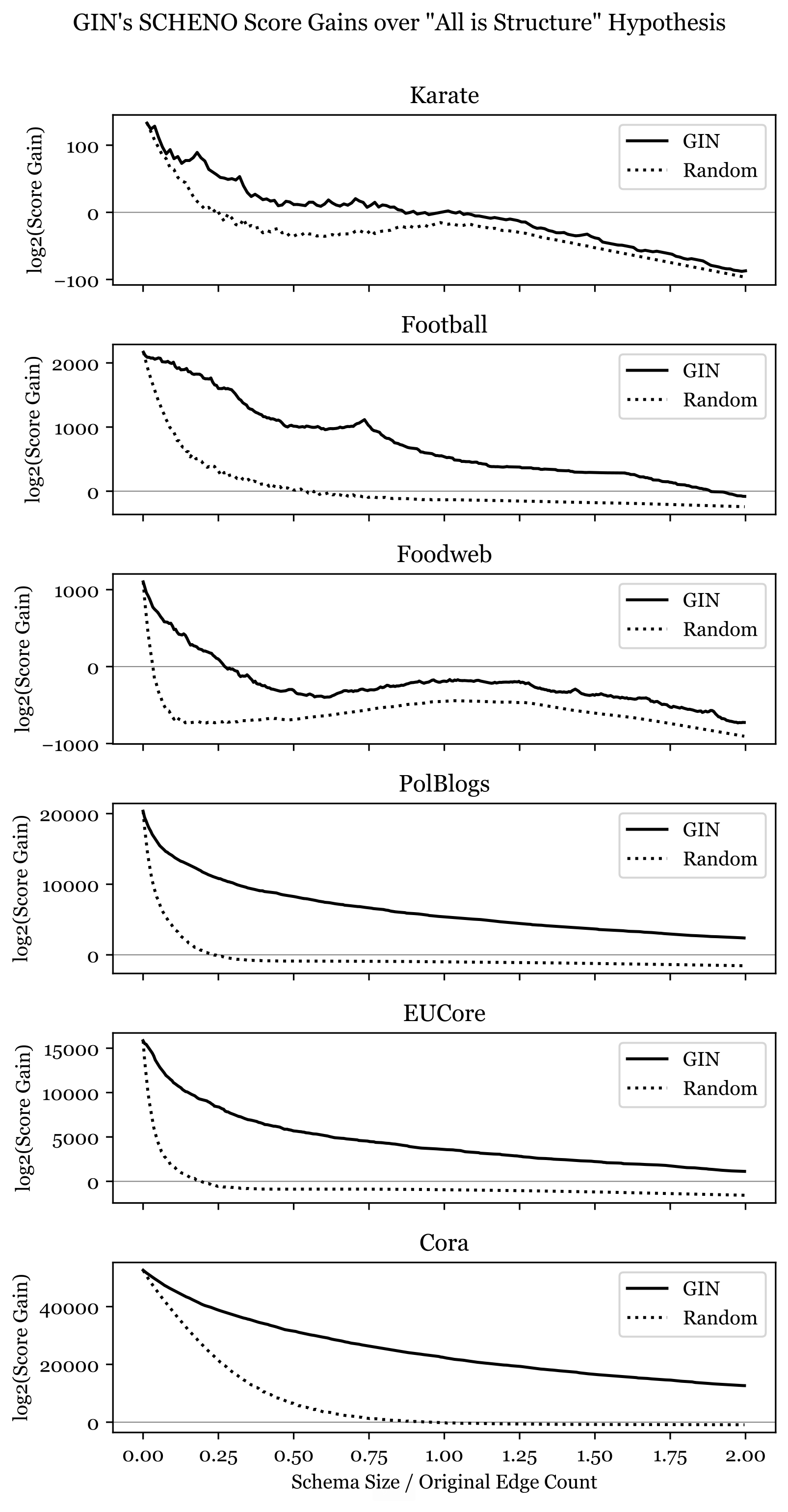}
\caption{\ntext{\textbf{Performance of Graph Isomorphism Network (GIN):} This figure shows how well a GIN does decomposing various graphs. The $y$-axis shows how much more likely SCHENO says GIN's decomposition is than the trivial decomposition (Schema = Graph, Noise = $\emptyset$). The $x$-axis expresses how many edges are in the schema as a fraction of the number of edges in the original graph; we show schemas ranging from 0 edges to 2$|E|$ edges. The ``Random'' decompositions are obtained by randomly adding and removing as many edges as the GIN added and removed from the original graph when the GIN obtained its schema. Note that GIN and Random always converge when the schema size is zero because there is only a single 0-edge schema (the empty graph).}}\label{fig:models_gin}
\end{figure}

\ntext{The results for GIN are shown in Figure~\ref{fig:models_gin}. Regardless of how many edges we require the GINs' schemas to contain, they consistently beat both the ``all is structure'' hypothesis and schemas created by randomly adding and/or deleting the same number of edges as the GIN. The key exception to this trend is the Foodweb network, in which the GIN does not find patterns which improve the score other than patterns akin to the ``no schema'' hypothesis obtained when deleting 75\% or more of the edges. We suspect that the reason the GIN performed poorly on the Foodweb network is similar to the reason that VoG struggled: the structure of the interlocking bipartite cores is more combinatorially complex than the triadic closure and degree distributions tendencies known to be present in the other networks we analyzed.}

\ntext{Finally we note that though GINs perform better with smaller schema sizes -- likely because SCHENO considers the original graph more noise than pattern -- they can still improve over the ``all is structure'' hypothesis by increasing rather than decreasing the schema size.}

\section{Directly Unearthing what SCHENO Prioritizes}\label{sec:algorithm}

\revonetext{Though SCHENO is meant as a metric of success for the schema-noise decomposition task rather than an algorithm, it can still be used to guide a search process. To get a better sense for what kinds of patterns the SCHENO function is capable of recognizing and prioritizing, we run a genetic algorithm to search for schema-noise decompositions with SCHENO as its objective (its fitness function). Though a simple genetic algorithm is not guaranteed to find whatever decompositions SCHENO considers the best, it will at least give us a rough perspective on what SCHENO incentivises. We call this genetic algorithm SCHENO GA.}

\subsection{Algorithmic Details for SCHENO and SCHENO GA}

Though many intricacies went into programming an efficient system, the basic idea is simple: Given a graph $G$, use a genetic algorithm to search for candidate noise sets. The fitness function is the derivative of SCHENO given in equation~\ref{eqn:score_practice} parametrized with a noise probability $p$ according to equation~\ref{eqn:final_p_formula}; see Section~\ref{sec:choosing_p} for more information on how we actually calculate a value for equation~\ref{eqn:final_p_formula}.

Recall that the data graph $G$ plus the noise set $N$ entails the hypothesis (\ie, schema) graph $H = G \oplus N$. To calculate the score for $(H, N)$, we need to make two distinct graph automorphism computations. For this we use \verb|traces|, which forms the computational bottleneck of our process~\cite{mckay2014practical}. \revonetext{Our codebase also includes the option to run the Weisfeiler Lehman (WL) algorithm instead of} \verb|traces|\revonetext{; when fully optimized, the WL algorithm can run in O$((n + m) \log n)$ time~\cite{berkholz2017tight}, and only a few calls to WL are needed to accurately approximate the total automorphism count.}
%, so our project is coded in \verb|C++|, wherein we call the \verb|C| code of the efficient automorphism program \verb|traces|
We parallelize the genetic algorithm's scoring process so that many automorphism calculations can happen simultaneously.

\subsection{Automorphism Calculation}

Calculating $|\Aut(H)|$ is fairly easy, as \verb|traces| is designed for that. Calculating $|\AO_H(N)|$ is trickier, as it is the automorphism orbit size of a set of edges and/or non-edges. We use the orbit-stablizer theorem as applied to graphs, which tells us that $|\AO_H(N)| = \frac{|\Aut(H)|}{|\text{Stab}_H(N)|}$~\cite{rose2009action}.

As a reminder, $\Stab_H(N)$ is the set of automorphisms in $H$ that map $N$ to itself. Fortunately for us, \verb|traces| permits node colors and will only find automorphisms that map nodes of the same color to each other. Thus, to find the stabilizers, we supplement the graph with nodes that correspond to edges. In an undirected graph, this means that an edge $(a, b)$ is converted into two edges $(a, c)$ and $(c, b)$ where node $c$ represents the edge; edge-nodes like $c$ are always given a different color from normal nodes like $a$ and $b$ so that \verb|traces| never treats edge-nodes like normal nodes. On a directed graph, we need two nodes per edge: $(a, b)$ becomes $(a, c_1)$, $(c_1, c_2)$ and $(c_2, b)$; the colors on nodes $c_1$ and $c_2$ indicate whether the edge goes from $a$ to $b$, $b$ to $a$, or both. For stabilizer purposes, we give the edge-nodes for added edges a color corresponding to \textit{addition}, and the edges that were removed a \textit{deleted} color. This means that when \verb|traces| runs on our augmented graph, it can only map deleted edges to deleted edges, added edges to added edges, and un-touched edges to un-touched edges, thereby giving us the stabilizer for the noise set.

%To make calls to \verb|traces| efficiently, we want to be able to edit $G$ with one noise set $N_1$, restore $G$ to its former self, edit $G$ with another noise set $N_2$, and so on, rather than creating an entire graph from scratch for each new noise set to be scored. Unfortunately, the graph representation that \verb|traces| and its partner program \verb|nauty| take as input is not amenable to edits, as all edges must be in an adjacency list stored in a single array. Thus part of making our code efficient was to find a way to edit this list quickly. We created something akin to a greatly-simplified malloc that allows edits to the graph in amortized constant time by adapting the logic from variable-length arrays. 
\revonetext{We provide pseudocode for the undirected case in Algorithm~\ref{alg:scheno_undirected} in the appendix. Note that this pseudocode merely shows the logic; our actual code is written much more efficiently and with numerical stability in mind.} Complete source code and documentation are available at \url{https://github.com/schemanoise/SCHENO}.

%Traces is also not designed to handle directed graphs, but our use of nodes to represent edges, combined with node colors to distinguish edge-nodes from normal nodes, allows us to include directionality.

\subsection{Genetic Algorithm Optimizer}

%For those who want to know the exact technical details, we offer more information here about our algorithm. Our code is also available on github at: \url{https://github.com/schemanoise/SCHENO}.

Our genetic algorithm operates by growing the existing population tenfold and then shrinking it back to the original size by keeping the highest scoring members. Notably, the original population members are considered part of the expanded population, which means that a member of a population can survive indefinitely if it scores high enough. This formulation makes the algorithm rather greedy. We use a large population size so that exploration may still occur.

A population member is a set of node pairs, which may be edges and/or non-edges. The population size is $P = (n + 400)^{1.4}$. This number was selected partly according to algorithmic performance and partly according to our patience; it has no particular theoretical significance. We then create $6P$ new population members by randomly mutating our original $P$ members; mutations swap out an existing node pair for a new one with $60\%$ probability, and add or remove a node pair with $20\%$ probability each.

After mutations, we create $3P$ new population members by randomly mating two of the current population (the original $P$ and/or the newly mutated $6P$). If a node pair is in both parents, it is automatically kept; if it is just one parent, it is kept with a $50\%$ chance.

\begin{figure}[h]
\centering
\resizebox{\linewidth}{!}{
    \begin{tabular}{cc}
        \subfloat[]{\begin{tikzpicture}[scale=1.0]
\node [textnode] at (2, 3.5) {\Large $\Rightarrow$};
    \node [node] at (0.0, 3.0) (a1) {};
    \node [node] at (0.0, 4.0) (b1) {};
    \node [node] at (1.0, 3.0) (c1) {};
    \node [node] at (1.0, 4.0) (d1) {};
    \node [node] at (3.0, 3.0) (a2) {};
    \node [node] at (3.0, 4.0) (b2) {};
    \node [node] at (4.0, 3.0) (c2) {};
    \node [node] at (4.0, 4.0) (d2) {};
    \draw [edge] (a1) to (b1);
    \draw [edge] (a1) to (c1);
    \draw [edge] (b1) to (d1);
    \draw [edge] (b1) to (c1);
    \draw [edge] (c1) to (d1);
    \draw [edge] (a2) to (b2);
    \draw [edge] (a2) to (c2);
    \draw [edge] (a2) to (d2);
    \draw [edge] (b2) to (d2);
    \draw [edge] (b2) to (c2);
    \draw [edge] (c2) to (d2);
\end{tikzpicture}} & \subfloat[]{\begin{tikzpicture}[scale=1.0]
\node [textnode] at (2.25, 0.75) {\Large $\Rightarrow$};
    \node [node] at (0.0, 0.0) (a1) {};
    \node [node] at (-0.5, 0.75) (b1) {};
    \node [node] at (0.0, 1.5) (c1) {};
    \node [node] at (1.0, 1.5) (d1) {};
    \node [node] at (1.5, 0.75) (e1) {};
    \node [node] at (1.0, 0.0) (f1) {};
    \node [node] at (3.5, 0.0) (a2) {};
    \node [node] at (3.0, 0.75) (b2) {};
    \node [node] at (3.5, 1.5) (c2) {};
    \node [node] at (4.5, 1.5) (d2) {};
    \node [node] at (5.0, 0.75) (e2) {};
    \node [node] at (4.5, 0.0) (f2) {};
    \draw [edge] (a1) to (b1);
    \draw [edge] (b1) to (c1);
    \draw [edge] (c1) to (d1);
    \draw [edge] (e1) to (f1);
    \draw [edge] (f1) to (a1);
    \draw [edge] (a2) to (b2);
    \draw [edge] (c2) to (d2);
    \draw [edge] (e2) to (f2);
\end{tikzpicture}} \\
        \subfloat[]{\begin{tikzpicture}[scale=1.0]
\node [textnode] at (2.125, 0.9) {\Large $\Rightarrow$};
    \node [node] at (0.0, 0.0) (a1) {};
    \node [node] at (-0.5, 0.7) (b1) {};
    \node [node] at (-0.25, 1.4) (c1) {};
    \node [node] at (0.5, 1.8) (d1) {};
    \node [node] at (1.25, 1.4) (e1) {};
    \node [node] at (1.5, 0.7) (f1) {};
    \node [node] at (1.0, 0.0) (g1) {};
    \node [node] at (3.25, 0.0) (a2) {};
    \node [node] at (2.75, 0.7) (b2) {};
    \node [node] at (3.0, 1.4) (c2) {};
    \node [node] at (3.75, 1.8) (d2) {};
    \node [node] at (4.5, 1.4) (e2) {};
    \node [node] at (4.75, 0.7) (f2) {};
    \node [node] at (4.25, 0.0) (g2) {};
    \draw [edge] (a1) to (b1);
    \draw [edge] (b1) to (c1);
    \draw [edge] (c1) to (d1);
    \draw [edge] (e1) to (f1);
    \draw [edge] (f1) to (g1);
    \draw [edge] (g1) to (a1);
    \draw [edge] (a2) to (b2);
    \draw [edge] (b2) to (c2);
    \draw [edge] (c2) to (d2);
    \draw [edge] (d2) to (e2);
    \draw [edge] (e2) to (f2);
    \draw [edge] (f2) to (g2);
    \draw [edge] (g2) to (a2);
\end{tikzpicture}} & \subfloat[]{\begin{tikzpicture}[scale=1.0]
\node [textnode] at (2.125, 0.9) {\Large $\Rightarrow$};
    \node [node] at (0.0, 0.0) (a1) {};
    \node [node] at (-0.5, 0.7) (b1) {};
    \node [node] at (-0.25, 1.4) (c1) {};
    \node [node] at (0.5, 1.8) (d1) {};
    \node [node] at (1.25, 1.4) (e1) {};
    \node [node] at (1.5, 0.7) (f1) {};
    \node [node] at (1.0, 0.0) (g1) {};
    \node [node] at (0.5, 0.85) (h1) {};
    \node [node] at (3.25, 0.0) (a2) {};
    \node [node] at (2.75, 0.7) (b2) {};
    \node [node] at (3.0, 1.4) (c2) {};
    \node [node] at (3.75, 1.8) (d2) {};
    \node [node] at (4.5, 1.4) (e2) {};
    \node [node] at (4.75, 0.7) (f2) {};
    \node [node] at (4.25, 0.0) (g2) {};
    \node [node] at (3.75, 0.85) (h2) {};
    \draw [edge] (a1) to (h1);
    \draw [edge] (b1) to (h1);
    \draw [edge] (c1) to (h1);
    \draw [edge] (d1) to (h1);
    \draw [edge] (e1) to (h1);
    \draw [edge] (g1) to (h1);
    \draw [edge] (f1) to (g1);
    \draw [edge] (b1) to (c1);
    \draw [edge] (h2) to (a2);
    \draw [edge] (h2) to (b2);
    \draw [edge] (h2) to (c2);
    \draw [edge] (h2) to (d2);
    \draw [edge] (h2) to (e2);
    \draw [edge] (h2) to (f2);
    \draw [edge] (h2) to (g2);
\end{tikzpicture}} \\
        \subfloat[]{\begin{tikzpicture}[scale=1.0]
\node [textnode] at (2.75, 0.875) {\Large $\Rightarrow$};
    \node [node] at (0.0, 0.0) (a1) {};
    \node [node] at (0.5, 0.75) (b1) {};
    \node [node] at (1.0, 0.0) (c1) {};
    \node [node] at (0.0, 1.0) (d1) {};
    \node [node] at (0.5, 1.75) (e1) {};
    \node [node] at (1.0, 1.0) (f1) {};
    \node [node] at (1.0, 0.5) (g1) {};
    \node [node] at (1.5, 1.25) (h1) {};
    \node [node] at (2.0, 0.5) (i1) {};
    \node [node] at (3.5, 0.0) (a2) {};
    \node [node] at (4.0, 0.75) (b2) {};
    \node [node] at (4.5, 0.0) (c2) {};
    \node [node] at (3.5, 1.0) (d2) {};
    \node [node] at (4.0, 1.75) (e2) {};
    \node [node] at (4.5, 1.0) (f2) {};
    \node [node] at (4.5, 0.5) (g2) {};
    \node [node] at (5.0, 1.25) (h2) {};
    \node [node] at (5.5, 0.5) (i2) {};
    \draw [edge] (a1) to (b1);
    \draw [edge] (b1) to (c1);
    \draw [edge] (c1) to (a1);
    \draw [edge] (d1) to (e1);
    \draw [edge] (e1) to (f1);
    \draw [edge] (f1) to (d1);
    \draw [edge] (g1) to (h1);
    \draw [edge] (i1) to (g1);
    \draw [edge] (a2) to (b2);
    \draw [edge] (b2) to (c2);
    \draw [edge] (c2) to (a2);
    \draw [edge] (d2) to (e2);
    \draw [edge] (e2) to (f2);
    \draw [edge] (f2) to (d2);
    \draw [edge] (g2) to (h2);
    \draw [edge] (h2) to (i2);
    \draw [edge] (i2) to (g2);
\end{tikzpicture}} & \subfloat[]{\begin{tikzpicture}[scale=1.0]
\node [textnode] at (2.5, 2.3) {\Large $\Rightarrow$};
    \node [node] at (1.0, 5.0) (a1) {};
    \node [node] at (0.4, 4.4) (b1) {};
    \node [node] at (1.6, 4.4) (c1) {};
    \node [node] at (1.0, 3.8) (d1) {};
    \node [node] at (1.0, 2.8) (e1) {};
    \node [node] at (0.0, 2.8) (f1) {};
    \node [node] at (2.0, 2.8) (g1) {};
    \node [node] at (1.0, 1.8) (h1) {};
    \node [node] at (1.0, 0.8) (i1) {};
    \node [node] at (0.4, 0.2) (j1) {};
    \node [node] at (1.6, 0.2) (k1) {};
    \node [node] at (4.0, 5.0) (a2) {};
    \node [node] at (3.4, 4.4) (b2) {};
    \node [node] at (4.6, 4.4) (c2) {};
    \node [node] at (4.0, 3.8) (d2) {};
    \node [node] at (4.0, 2.8) (e2) {};
    \node [node] at (3.4, 3.4) (f2) {};
    \node [node] at (4.6, 3.4) (g2) {};
    \node [node] at (4.0, 1.8) (h2) {};
    \node [node] at (4.0, 0.8) (i2) {};
    \node [node] at (3.4, 0.2) (j2) {};
    \node [node] at (4.6, 0.2) (k2) {};
    \draw [edge] (a1) to (b1);
    \draw [edge] (a1) to (c1);
    \draw [edge] (b1) to (d1);
    \draw [edge] (c1) to (d1);
    \draw [edge] (d1) to (e1);
    \draw [edge] (f1) to (e1);
    \draw [edge] (g1) to (e1);
    \draw [edge] (e1) to (h1);
    \draw [edge] (h1) to (i1);
    \draw [edge] (i1) to (j1);
    \draw [edge] (i1) to (k1);
    \draw [edge] (d1) to (f1);
    \draw [edge] (c1) to (k1);
    \draw [edge] (a2) to (b2);
    \draw [edge] (a2) to (c2);
    \draw [edge] (b2) to (d2);
    \draw [edge] (c2) to (d2);
    \draw [edge] (f2) to (e2);
    \draw [edge] (g2) to (e2);
    \draw [edge] (e2) to (h2);
    \draw [edge] (h2) to (i2);
    \draw [edge] (i2) to (j2);
    \draw [edge] (i2) to (k2);
\end{tikzpicture}} \\
    \end{tabular}
}
\caption{\textbf{Transformations of Small Graphs} -- The results shown are mostly meant to be intuitive, but some (b and f) took us by surprise, and we leave them here. In the case of (f), our instinct to see human figures blinded us to the actual highly-symmetric structures nearby, structures that SCHENO GA finds.}\label{fig:simple_figs}
\end{figure}
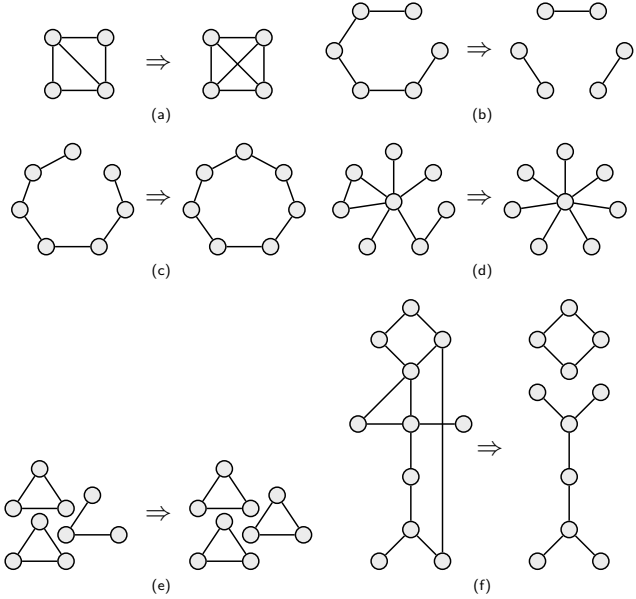

\begin{figure}[t]
\centering

\subfloat[Original Graph]{\includegraphics[width=.48\linewidth]{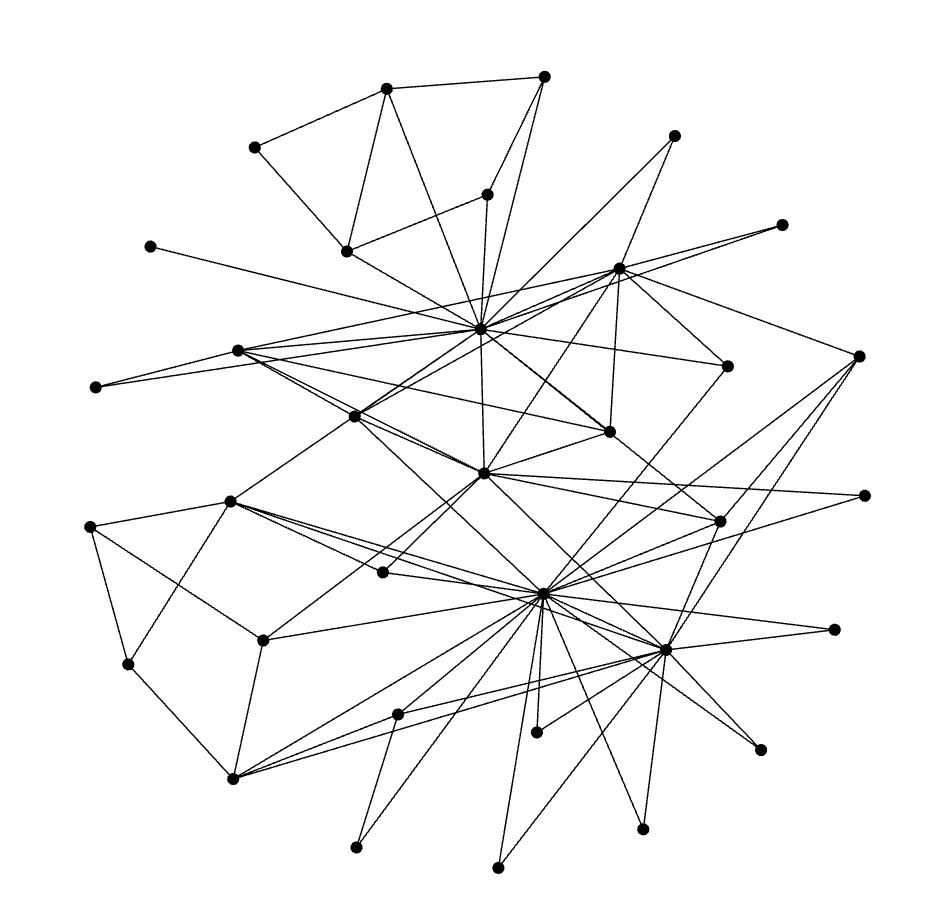}}
\subfloat[Schema-Noise Decomposition]{\includegraphics[width=.48\linewidth]{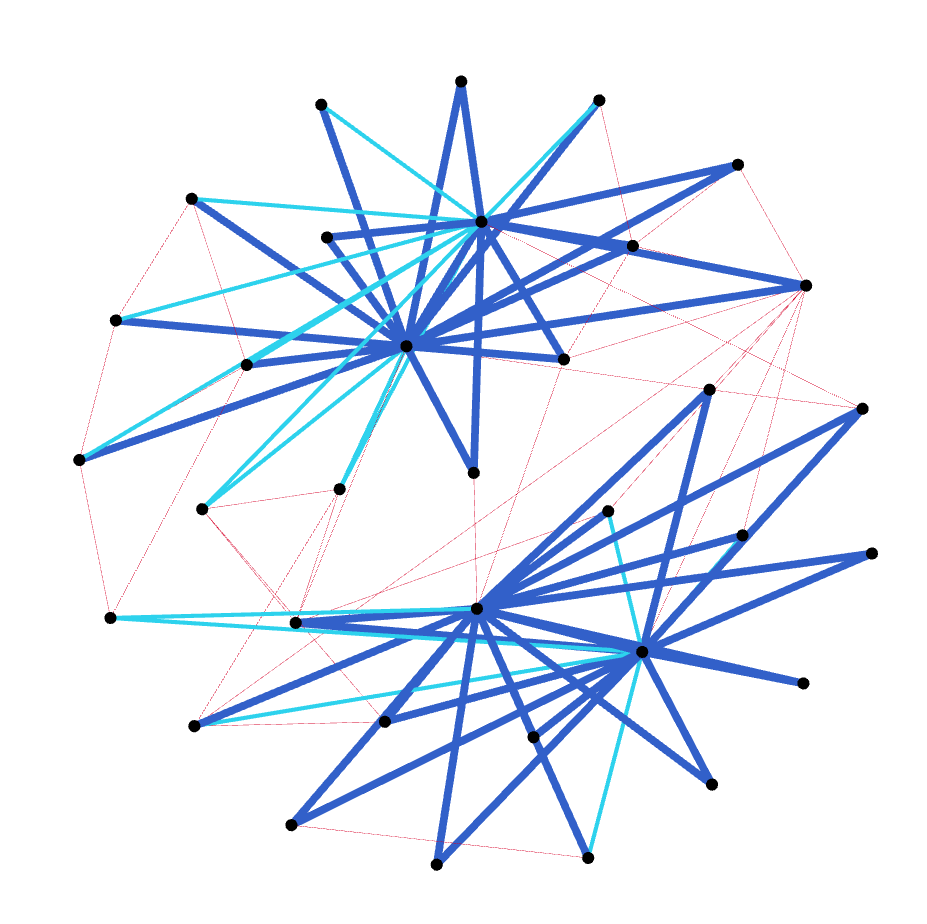}} \\
\subfloat[Schema Only]{\includegraphics[width=.30\linewidth]{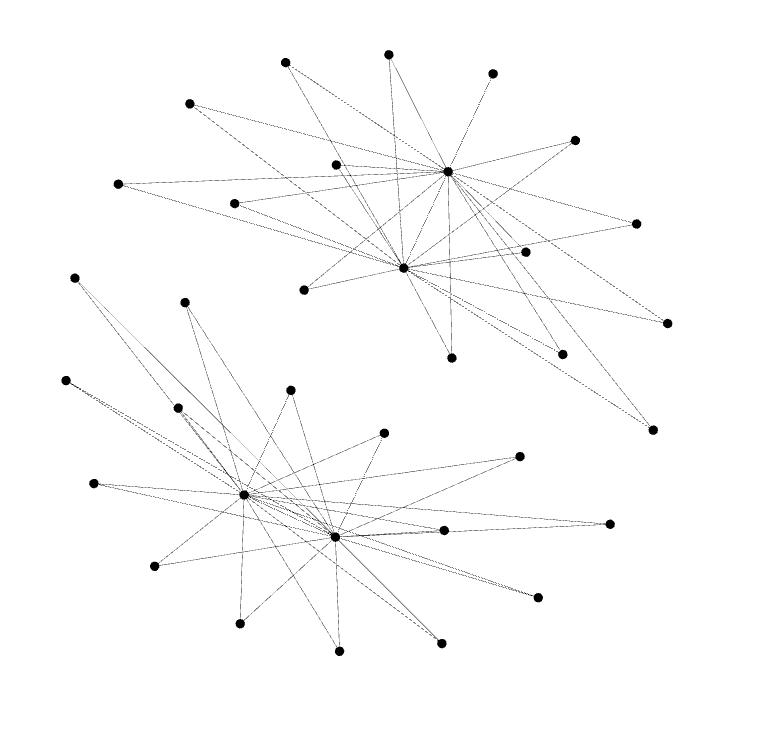}}
\subfloat[Noise Only]{\includegraphics[width=.30\linewidth]{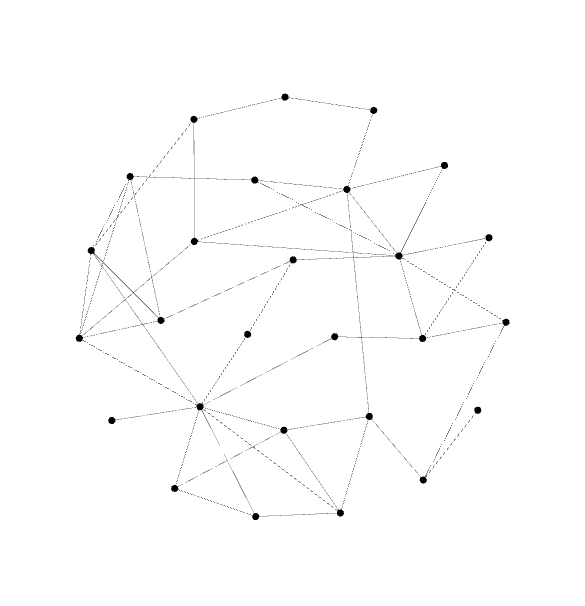}}
\caption{\textbf{Performance on the Karate Graph} -- Note: The node-layouts in the sub-figures are not identical; rather they are arranged individually for visibility. In \textbf{(b)}, dark blue represents an original edge, light blue represents an edge SCHENO GA added, and red represents one it deleted.}\label{fig:karate}
\end{figure}

\section{SCHENO GA Results}\label{sec:experiments}

We run SCHENO GA on five different kinds of graphs.

\begin{enumerate}
    \item Small, intuitive patterns corrupted by some noise
    \item Completely random graphs (to make sure SCHENO does not make something out of nothing)
    \item Highly-structured synthetic graphs corrupted by some noise
    \item Some real-world graphs
    \item Out of curiosity, we treat the black-and-white MNIST digit images as adjacency matrices for directed graphs and see how SCHENO GA modifies the images.
\end{enumerate}

In all of these experiments, the key question is not: can the genetic algorithm can find the optimal result according to SCHENO? for we know that genetic algorithms are limited. Rather, the questions are: does SCHENO reward actual patterns? and how many different \emph{kinds} of pattern can SCHENO incentivize?

Experiments 2, 3, and 5 are discussed in the appendix. We discuss experiments 1 and 4 in Sections~\ref{sec:small_patterns} and~\ref{sec:ga_real_world} respectively.

\subsection{Small, Intuitive Patterns}\label{sec:small_patterns}

On small examples, SCHENO GA successfully manages to \textit{fill in the gaps} and \textit{erase the noise} in a way that largely corresponds to human intuition. When the result does not correspond to initial expectations, it still makes intuitive sense. See Figure~\ref{fig:simple_figs} for examples.

\subsection{Real-World Data}\label{sec:ga_real_world}

\revonetext{When testing on real-world datasets, our goal is not for SCHENO GA to beat the various landmark graph-mining algorithms by getting higher SCHENO scores; SCHENO GA easily accomplishes that when given enough time and space. Rather, the goal is to empirically observe whether the principled assumptions we made in designing SCHENO actually translate into selecting genuine patterns in practice. In other words, do SCHENO scores actually differentiate between good and bad decompositions? SCHENO is our proposed metric for success, and we want to qualitatively determine if it corresponds well to the notions of structure and noise. It is logically possible that though \textit{SCHENO} might differentiate between good and bad decompositions as we want it to, \textit{SCHENO GA} might fail to find what SCHENO considers most ideal; thus our tests could fail to show SCHENO's full capacity. Fortunately, SCHENO GA still manages to uncover a wide variety of interesting patterns.}

Our most striking results are probably those we obtained on the smaller real-world datasets. We ran SCHENO GA on 6 real-world networks: Zachary's Karate Club network~\cite{konect}, an ecological food network (what eats what) from Little Rock Lake in Wisconsin~\cite{konect}, a 2004 college football network documenting which FBS teams played at least one game against another team~\cite{collegefootball}, the EU-Core email network~\cite{snapnets}, the Cora ML citation network~\cite{networkrepository}, and a political blogs network showing which blogs reference other blogs~\cite{konect}.

When running on the karate club network, SCHENO GA uncovers the famous historical schism. The network represents social relationships within a club. Eventually the club split into two clubs following two different headmasters. Our algorithm's decomposition of the graph into schema and noise can be seen in Figure~\ref{fig:karate}. Not only does SCHENO GA notice the split into two groups, but in different trials, our schema decomposition gets between 88 and 94\% of the memberships in the eventual historical split correct. Furthermore, we uncover an interesting feature of the graph: For both of the two groups, there is a clear central figure, but there is also a secondary figure that connects to many of the peripheral members of the group.

\begin{figure}[ht]
\centering
%\adjustbox{max width=0.8\linewidth}{%
\subfloat[Edges Added/Removed]{
\includegraphics[width=.48\linewidth]{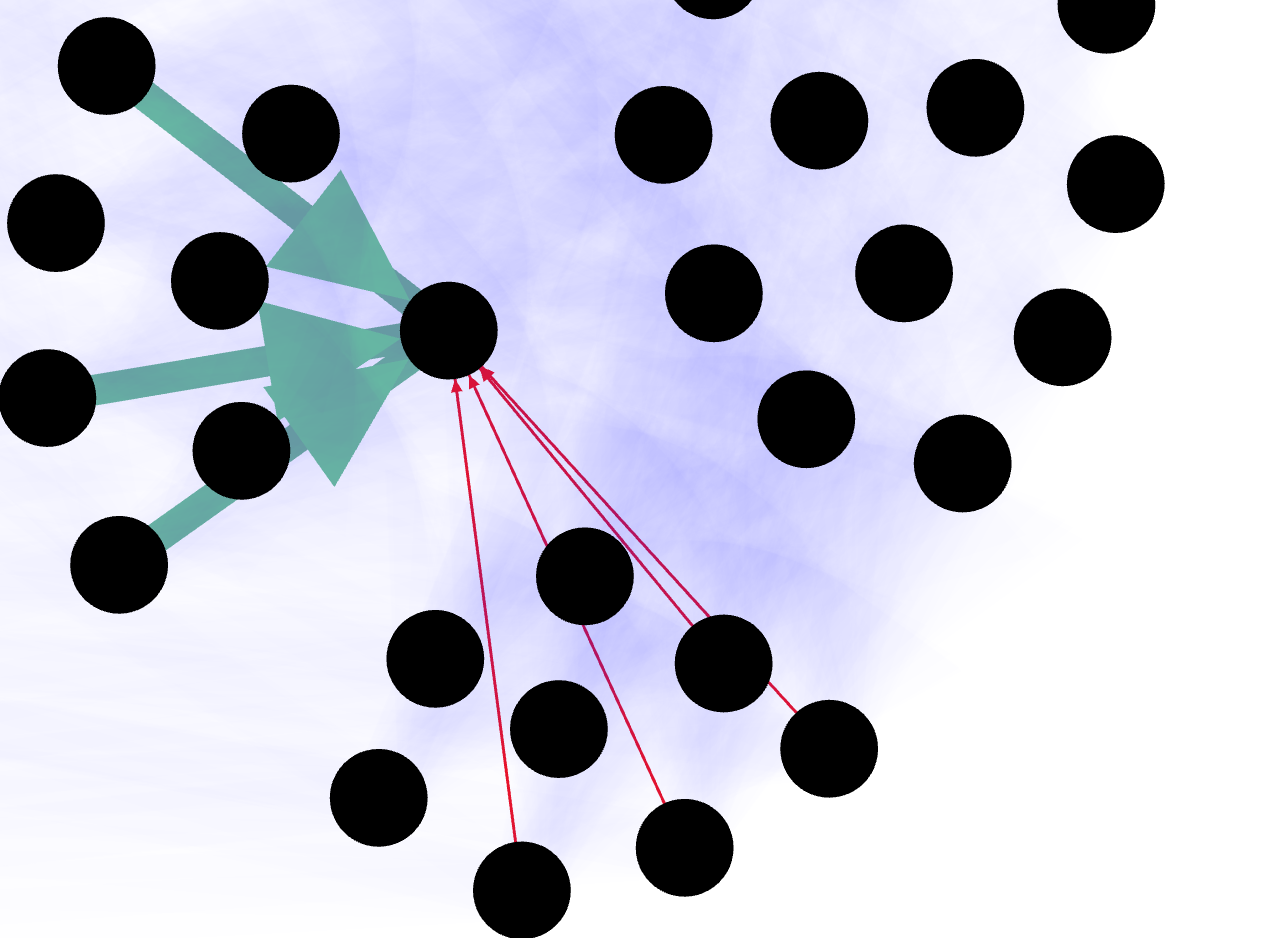}}
\subfloat[Schema Version with Full Symmetries]{\includegraphics[width=.48\linewidth]{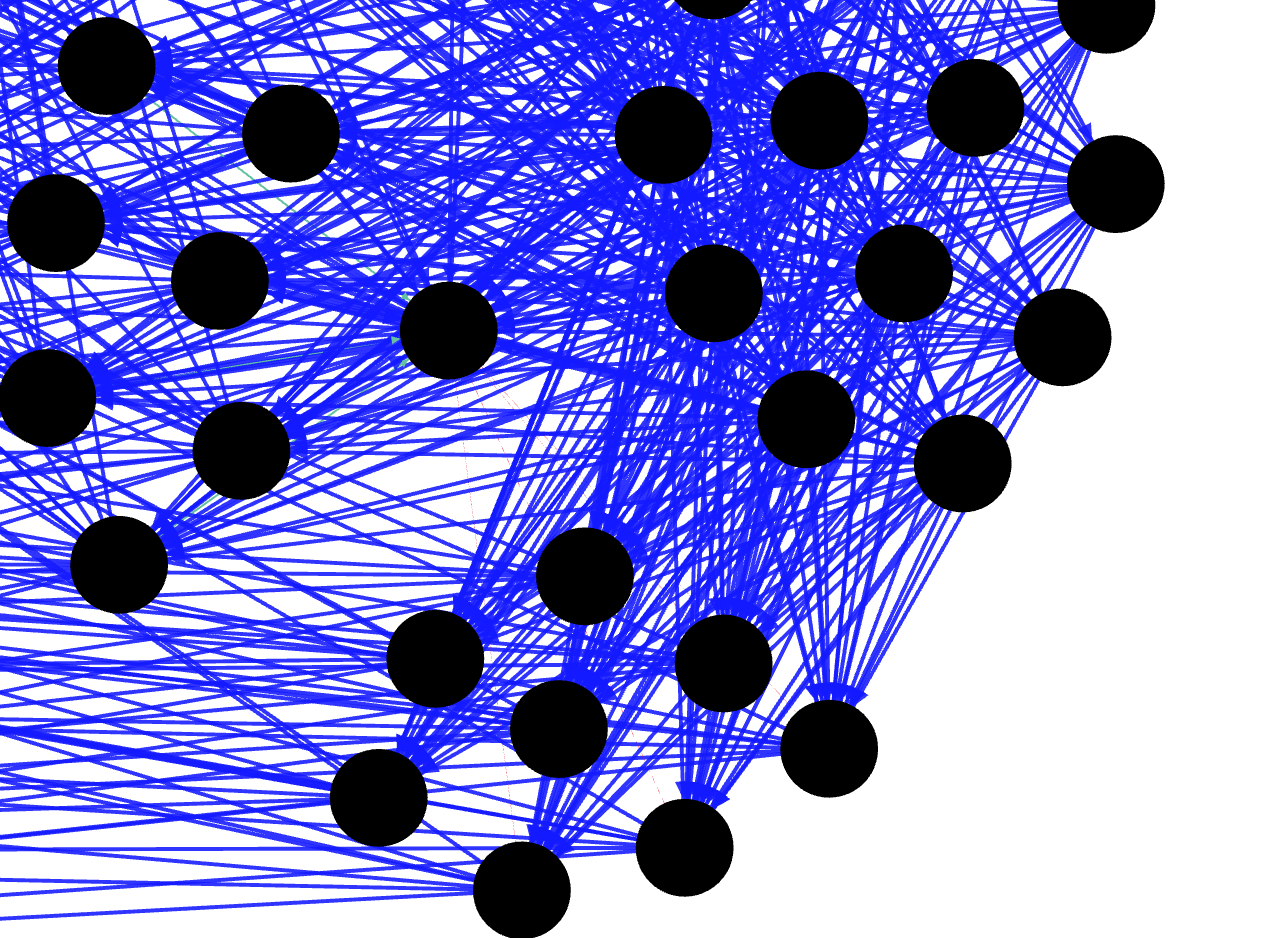}} \\
\subfloat[Edges Added/Removed]{\includegraphics[width=.48\linewidth]{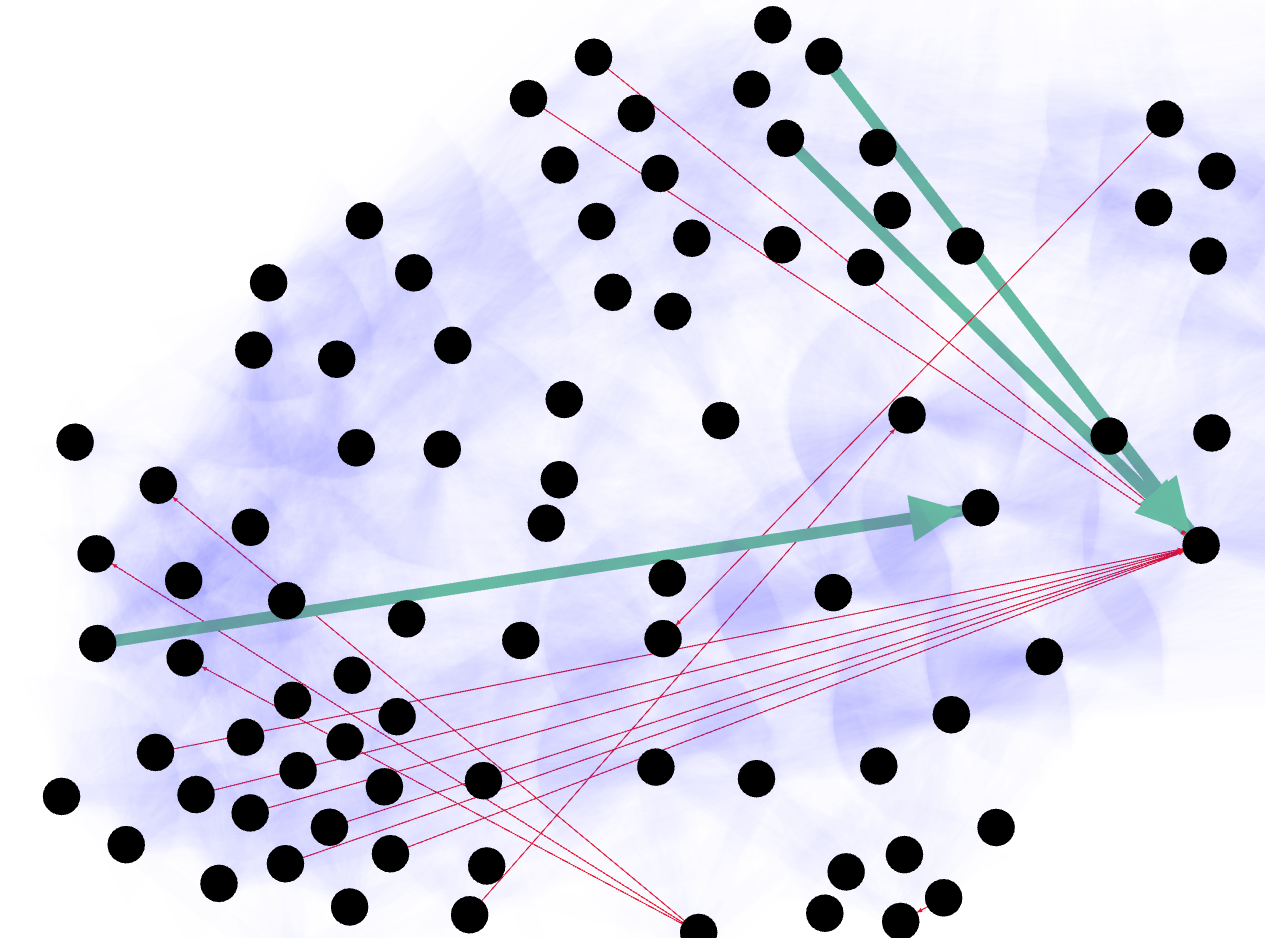}}
\subfloat[Schema Version with Full Symmetries]{\includegraphics[width=.48\linewidth]{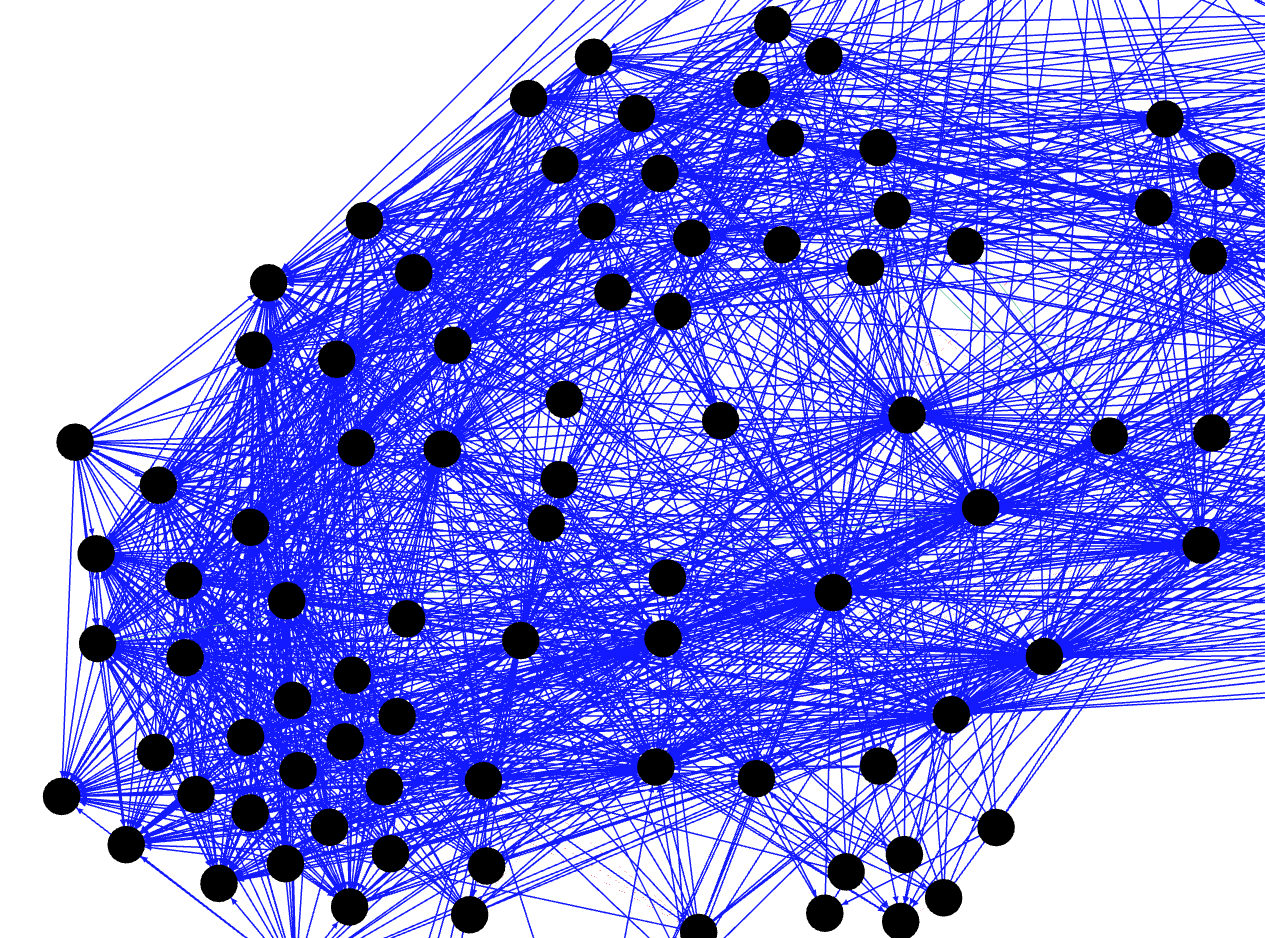}}
%}
\caption{\textbf{Performance on the Moreno Foodweb} -- SCHENO GA finds several anomalous species that eat and/or are eaten by other species in an irregular way. By rectifying these anomalies, the general pattern is restored to the graph. Teal means added; red means deleted.}\label{fig:foodweb}
\end{figure}

On the ecological food network, SCHENO GA uncovers several anomalies. Broadly speaking, in the network there are groups of species that eat the same things and are eaten by the same things (\eg, a group of bugs that all eat the same plants and are eaten by the same birds). However, anomalies disrupt these symmetries, and our algorithm picks up on this fact. See Figure~\ref{fig:foodweb} for illustrations.%We were curious to know which real-life species these anomalous nodes corresponded to, but unfortunately were not able to find that data. 
The results on this graph suggest that in some cases SCHENO-based structure-finding can coincide with anomaly-detection.

\begin{figure}[ht]
\centering
\subfloat[Connections Removed]{\includegraphics[width=.48\linewidth]{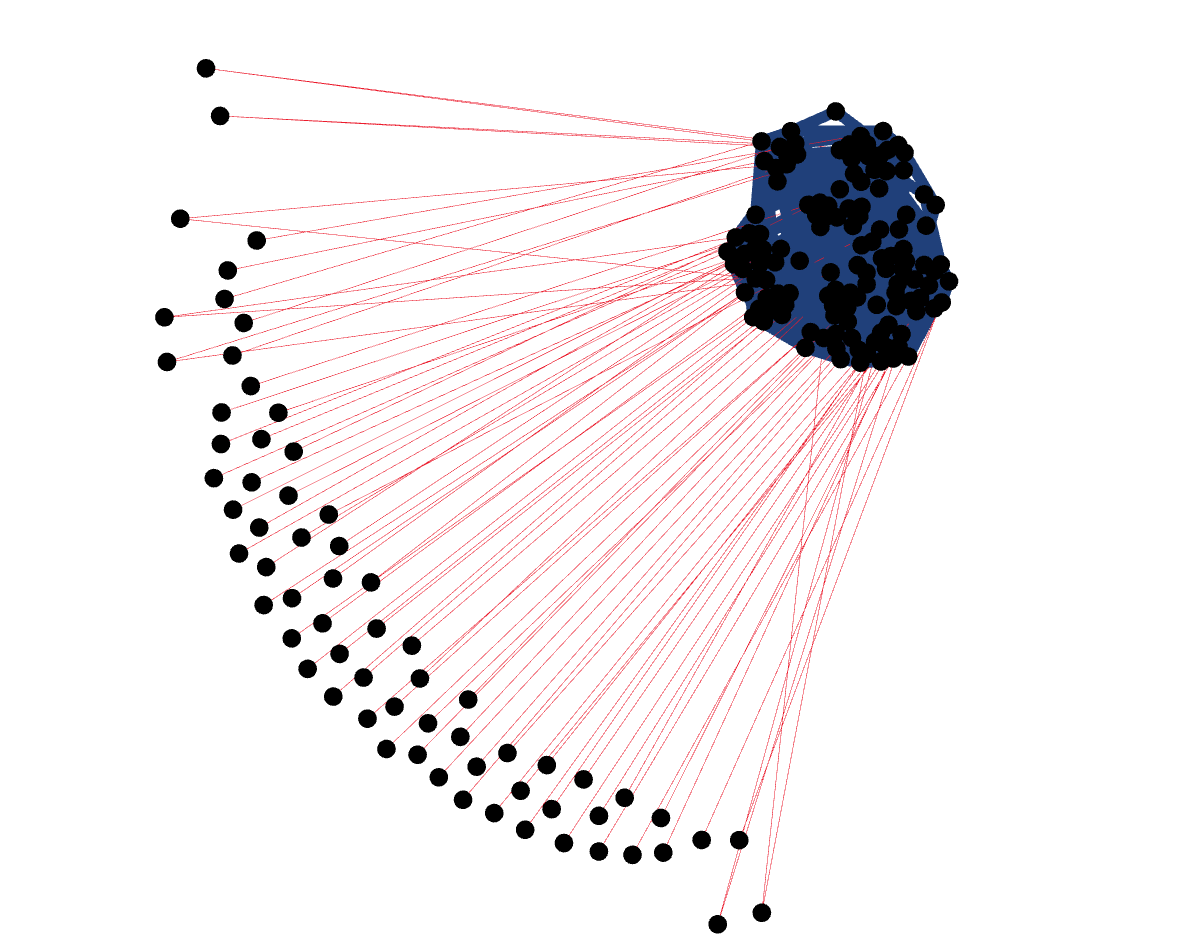}}
\subfloat[Zoomed-in View of Mostly-FBS Games]{\includegraphics[width=.48\linewidth]{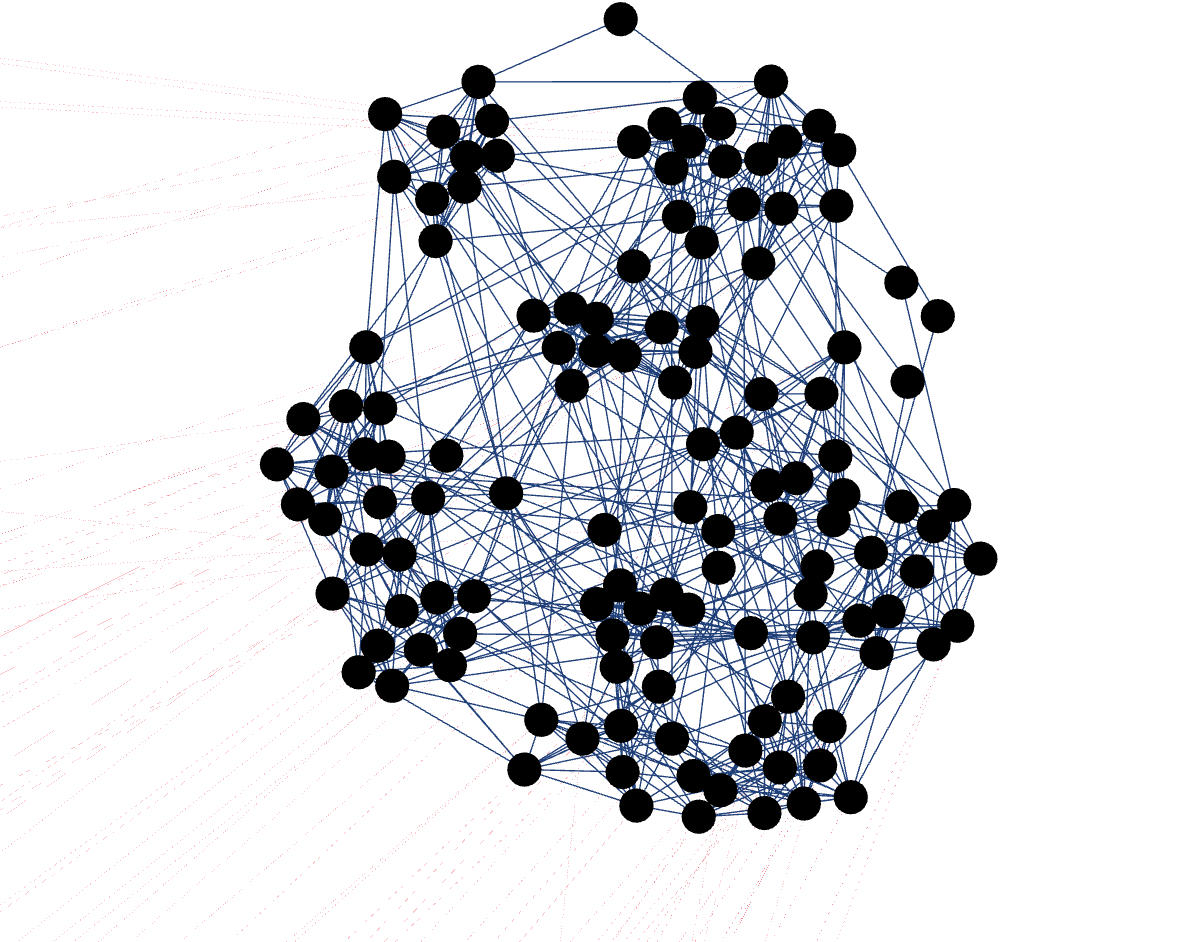}}
\caption{\textbf{Performance on a College Football Graph} -- The non-FBS teams (aka FCS) get disconnected from the core network.}\label{fig:football}
\end{figure}

The college football dataset connects two teams if they played at least one game in the 2004 season. The dataset is limited to games with at least one FBS team. SCHENO GA's one finding was to delete all the edges to non-FBS teams (FCS), treating the schema as the FBS games and the noise as the few edges to FCS teams. This results in a symmetry gain because the newly-disconnected FCS teams all become interchangeable with each other. We think this split makes sense, because from a structural perspective, choosing which FCS team an FBS team plays seems arbitrary. We show the results in Figure~\ref{fig:football}. We suspect that if SCHENO GA was able to better-explore the search space, it might find further adjustments within the FBS-only games as well, but the relatively greedy process makes no progress on that cluster even though SCHENO might better-reward another solution.

\begin{figure}[ht]
\centering
\subfloat[Some Modifications]{\includegraphics[width=.48\linewidth]{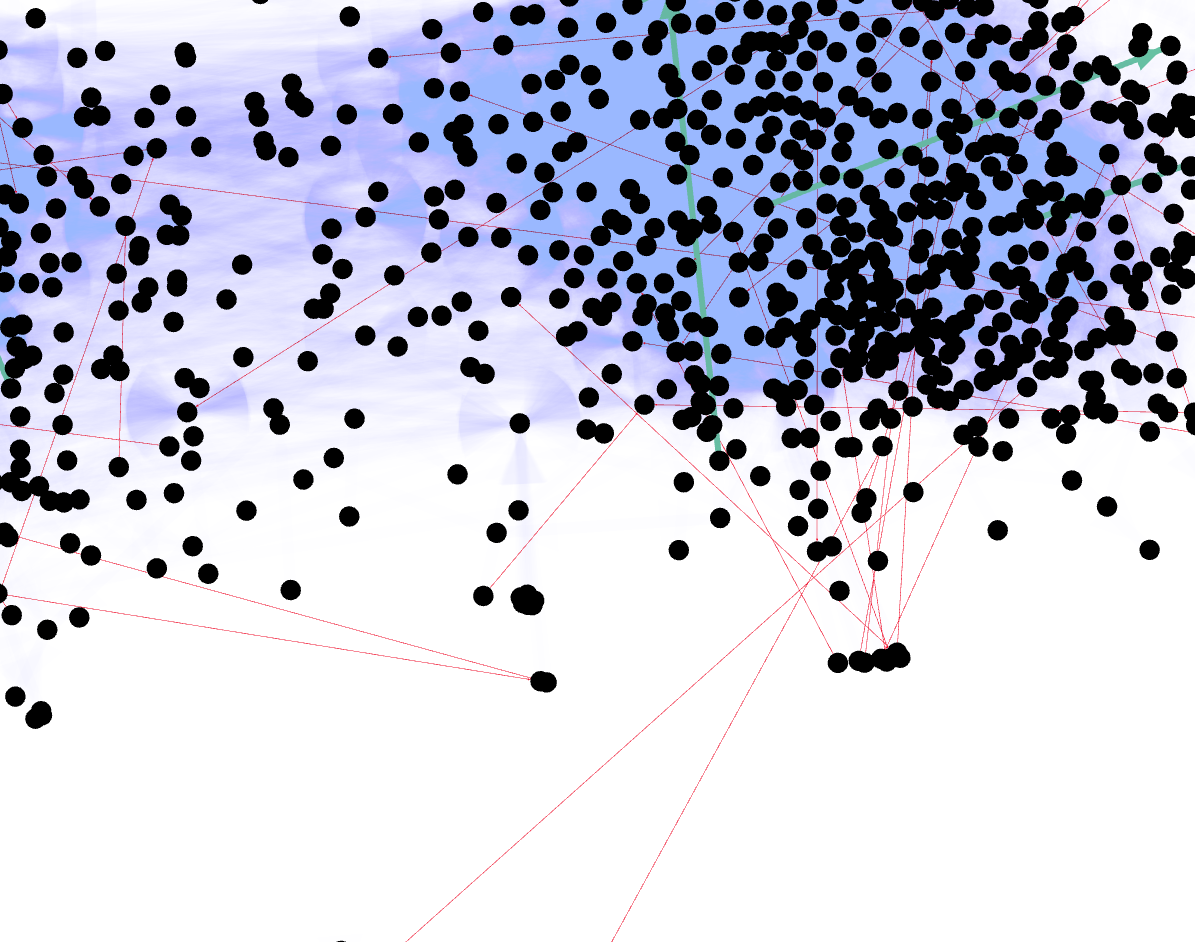}}
\subfloat[Resulting Structure]{\includegraphics[width=.48\linewidth]{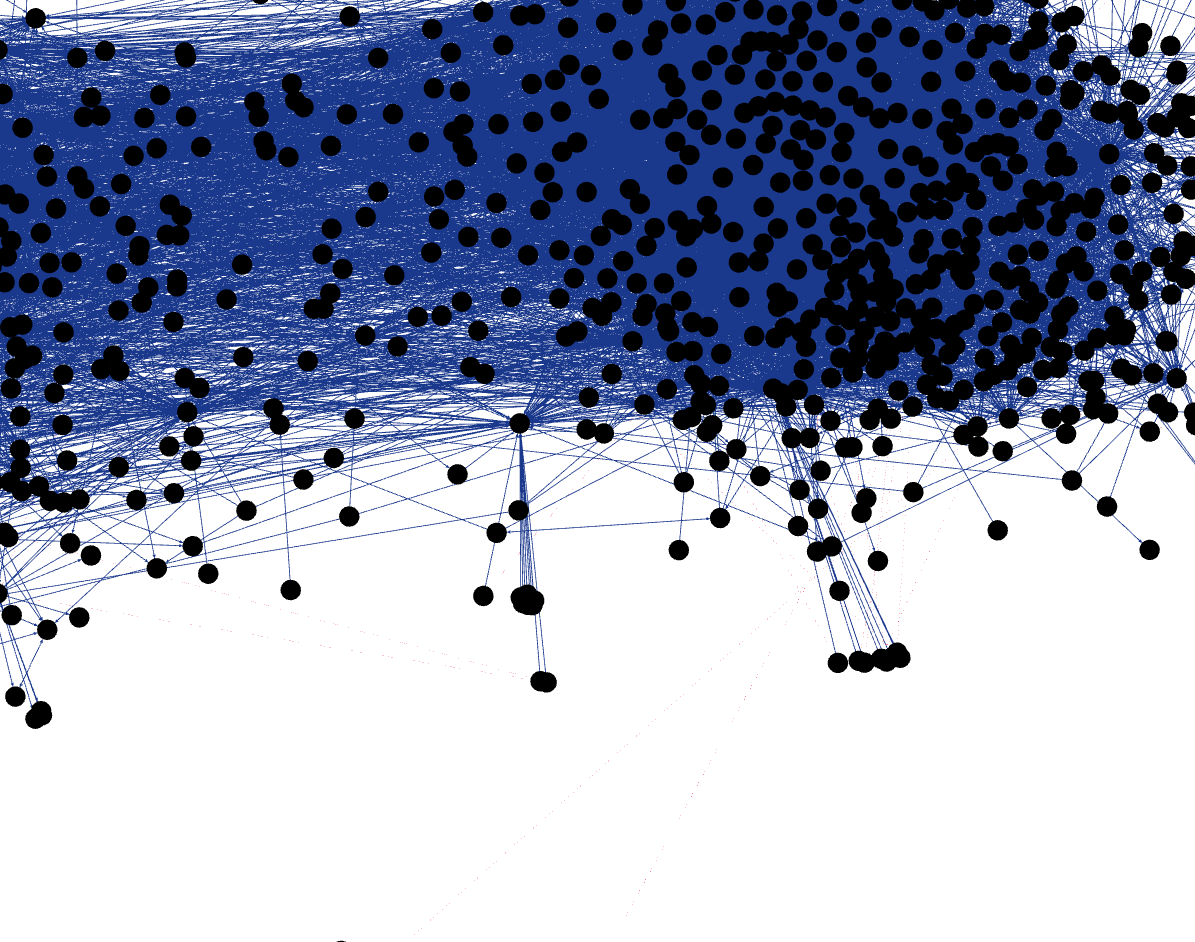}}
\caption{\textbf{Performance on Political Blogs Network} -- The change we highlight here (bottom-right of subfigures) shows the algorithm finding a cluster of nodes that each cite one particular blog and not much else. Again, teal means an edge is added; red means it is deleted.}\label{fig:pol_blogs}
\end{figure}

The remaining three graphs, Political Blogs, EU-Core Emails, and Cora Citations receive a few minor adjustments from SCHENO GA, but they are not as interesting. As in the college football graph, our algorithm mostly just deletes connections to nodes with few connections and then leaves a dense and mostly rigid core. There are a few exceptions to this in the political blogs network; we highlight one in Figure~\ref{fig:pol_blogs}. As with the football graph, we suspect that this limit on results is due to a lack of power in our search algorithm rather than a lack of power of the SCHENO scoring function, though at the moment we have no certain evidence one way or the other.

\ntext{The runtime of SCHENO GA varied from minutes to days depending on the size of the graph.}

In summary, SCHENO GA finds a wide variety of phenomenon across real-world data. The overall trend on real-world data seems to be that the patterns our genetic algorithm can most easily find are those where a group of nodes almost all have identical neighborhoods. Whether or not it is worth making nodes have identical neighborhoods is determined by SCHENO's specifications for the cost of editing the graph versus the reward of symmetry gains.

\section{Concluding Discussion}\label{sec:discussion}

The SCHENO scoring function offers a principled, goal-agnostic way of measuring how good a job one has done splitting apart a graph into schema and noise. It indicates that some well known graph-mining models have gaps---graphs wherein the patterns they find do not represent the graph's overall structure and/or do not represent the graph any better than a randomly selected pattern of the same size.

When even a relatively simple process like a genetic algorithm uses SCHENO to produce schema-noise decompositions, we uncover a wide variety of patterns in a wide variety of cases, from intuitive small patterns, to real-world graphs, to combinatoric structures and even image data (see appendix). This demonstrates that SCHENO is a general metric.

%SCHENO GA never produces a false positive; that is, it never uncovers a schema that's not truly a pattern, nor does it ever uncover a schema that does not closely relate to the original data.

Given the principled origins of our metric, we strongly suspect that with better, more sophisticated search algorithms guided by SCHENO, even more fascinating decompositions can be unearthed on larger and more complex graphs. Another step for future research would be to generalize SCHENO to graphs with node and edge types, or to graphs with weighted edges. Therefore, we expect that these definitions and formulae will likely prove useful to scientists in other fields attempting to uncover new patterns in structured data.

\section*{Acknowledgements} We would like to thank Daniel Gonzalez Cedre. This research was supported by the US National Science Foundation (\#1652492).

\bibliographystyle{plain}  % Need a bibliographystyle for elsarticle
\bibliography{references}

\appendix

\subsection{Algorithm Pseudocode}
\label{sec:alg}

Pseudocode on the undirected version of SCHENO can be found in Alg.~\ref{alg:scheno_undirected}.

% \begin{algorithm} %[H]
\customalg
\caption{Undirected Version of SCHENO}\label{alg:scheno_undirected}
\begin{algorithmic}

\Procedure{\\ \hfill SCHENO}{} ($V$: vertices, $E$: graph edges, $N$: noise edges)
\State $p \gets$ PARAM-CHOICE($|V|$)
\State schemaEdges $\gets E \oplus N$
\State schemaAutCount $\gets$ TRACES($V$, schemaEdges)
\State allEdges $\gets E \cup N$
\State addedEdges $\gets N \setminus \text{schemaEdges}$
\State deletedEdges $\gets N \cap \text{schemaEdges}$
\State noiseOrbitSize $\gets$ \\ \hfill NOISE-ORBIT-SIZE($V$, schemaAutCount, allEdges, \\ \hfill addedEdges, deletedEdges) \\ \vspace{-3mm}
\State \Return $\log_2($schemaAutCount$) + \log_2($noiseOrbitSize$)$ \\ \hspace{22mm} $+ |N|\log_2(p) + (\binom{|V|}{2} - |N|)\log_2(1 - p)$
\EndProcedure \\

\Procedure{PARAM-CHOICE}{$n$}
\If{$\langle n \text{ is small enough} \rangle$} \\ \Comment{See Section~\ref{sec:choosing_p} for details \ \ \ \ \ \ \ \ \ \ \ \ \ \ \ \ \ \ \ \ \ \ \ \ \ \ }
    \State sumAut $\gets \langle \text{exact, pre-computed value} \rangle$
\Else \\ \Comment{Redundancies left here to make connection to \ \ \ \ \  \\ \hfill formulae clearer \ \ \ \ } \\ \vspace{-3mm}
    \State numGraphs $\gets \frac{2^{\binom{n}{2}}}{n!}$ \\ \hfill $\cdot \left[1 + \frac{n(n-1)}{2^{n-1}} + \frac{(n!)}{(n-4)!}\frac{(3n-7)/(3n-9)}{2^{2n-3}} + \frac{n^5}{2^{5n/2}} \right]$ \\ \vspace{-1mm}
    \State sumAut $\gets \frac{(n!)(\text{numGraphs}^2)}{2^{\binom{n}{2}}}$
\EndIf
\State \Return $1 - \sqrt[\binom{n}{2}]{\frac{n!}{\text{sumAut}}}$
\EndProcedure \\

\Procedure{\\ \hfill NOISE-ORBIT-SIZE}{} ($V$, schemaAutCount, allEdges, \\ \hfill addedEdges, deletedEdges) \\ \vspace{-4mm}
\State stabNodes $\gets V\ \cup$ allEdges \Comment{For convenience, some \\ \hfill nodes here are ints and \\ \hfill others are int pairs} \\ \vspace{-4mm}
\State stabEdges $\gets \{(a, (a, b))\ |\ (a, b) \in \text{allEdges}\}$ \\ \hspace{20.5mm} $\cup \  \{(b, \hspace{0.3mm} (a, b))\ |\ (a, b) \in \text{allEdges}\}$
\State nodePartition $\gets \langle \text{empty hash map} \rangle$
\For{$v \in \text{stabNodes}$}
    \State nodePartition[$v$] $\gets$ \\ \vspace{1mm} \hfill $\begin{cases} \text{NODE} & v \in V \\ \text{ADDED EDGE} & v \in \text{addedEdges} \\
	     				           \text{DELETED EDGE} & v \in \text{deletedEdges} \\ \text{UNMODIFIED EDGE} & \text{otherwise} \end{cases}$
\EndFor
\State stabilizerAutCount $\gets$ \\ \hfill TRACES(stabNodes, stabEdges, nodePartition) \vspace{1mm}
\State \Return $\frac{\text{schemaAutCount}}{\text{stabilizerAutCount}}$
\EndProcedure
\end{algorithmic}
\end{algorithm}

\subsection{More SCHENO GA Results}

\subsubsection{Running on Total Noise}

As a sanity check, we generate some Erd\H os-R\'enyi graphs with edge probability $\frac{1}{2}$, then see what our code finds. We test on 50, 100, 150, and 200 nodes, 3 trials each. As we expected, SCHENO GA does not report a decomposition into structure and noise; it finds nothing, as there is nothing to find.

Technically, this means that SCHENO GA reports that $H = E$ and $N = \emptyset$ (all schema, no noise). We would like this to be reversed, but we expected it is because our genetic algorithm is greedy and would have to delete \emph{many} edges from $G$ before it would start finding that the whole graph was noise. In absence of finding that the entire graph is noise, we think that finding nothing is best, and that is what SCHENO seems to incentivise.

The more crucial question is: Which decomposition would get a better SCHENO score if the algorithm \emph{did} manage to find it? This was where we found one surprise: our model's noise probability $p$ is far enough from the $\frac{1}{2}$ probability we used to generate the noise graphs, that the noise is considered more likely to be structure simply due to the number of edges. This means there is a limit on the applicability of SCHENO: If graphs have many more edges than the proportion of possible edges dictated by our noise probability $p$, even noise might start to look a bit like structure.

Fortunately, this is not much of a problem for SCHENO for four reasons: 1) $p$ gets very close to $\frac{1}{2}$ (\eg\ when $n \geq 300$, $p > 0.467$). 2) Anyone wishing to extract schema from a graph with over half the possible edges can run on the complement of the graph. 3) People are unlikely to apply SCHENO to random noise, and even less likely to do so on random noise with proportion of edges $> p$. 4) If we generate the Erd\H os-R\'enyi graphs using edge probability $p$, then SCHENO correctly says that the entire graph is better explained as noise than as structure (and SCHENO GA still reports to have found nothing).

\subsubsection{Synthetic Structures}

Here we considered four combinatoric structures: a 128-node ring, a 127-node balanced binary tree, a 128-node wreath (like a ring, but each node connects to 4 neighbors on each side rather than 1 on each side), and the (10, 3) Johnson graph, which has 120 nodes. An $(a, b)$ Johnson graph has $\binom{a}{b}$ nodes, where each node represents and $a$-size subset of a set of $b$ elements; two nodes are connected if their corresponding sets differ in regard to only one element. Johnson graphs have a high degree of structure, and in fact are fundamentally related to the computational difficulty of the graph isomorphism problem~\cite{babai2016graph}.

For each graph, we randomly add and remove a small percentage of edges, then check two things: (1) Would SCHENO incentivise ``fixing'' the structure? (2) What does SCHENO GA find in practice. These synthetic tests were probably our most lackluster results, but they are still interesting to explore.

Our largest surprise was that the ring and binary tree were too sparse to be considered structure. SCHENO GA effectively said, ``There are so few edges that this is basically the empty graph.'' The empty graph is not an interesting schema from a human perspective, but it is highly structured in the sense that it has $n!$ automorphisms (each node is interchangeable with any other); in this sense it is equivalent to its complement: the complete graph.

We are not sure at present whether this result is a bug or a feature of SCHENO. On the one hand, the resulting pattern is not particularly fascinating. On the other hand, we tried to define schema in a principled way, and given our conceptual, philosophical setup, this might be the logically ``correct'' result. Furthermore, SCHENO says that the original synthetic structure is a good schema -- much more so than the perturbed graph; SCHENO just says that listing the whole perturbed graph as noise is an even \emph{better} decomposition.

The wreath graph was interesting, because rather than recover the original wreath (modified slightly by noise), SCHENO GA surprised us and pointed out many local symmetries. It found that with a few edits, it could make two or three neighbors in the wreath interchangeable. Thus the algorithm converted the graph into something like a ring lattice. See Figure~\ref{fig:wreath} for some visualizations.

\customfig
\centering
\subfloat[Less Original Noise]{\includegraphics[width=0.264\linewidth]{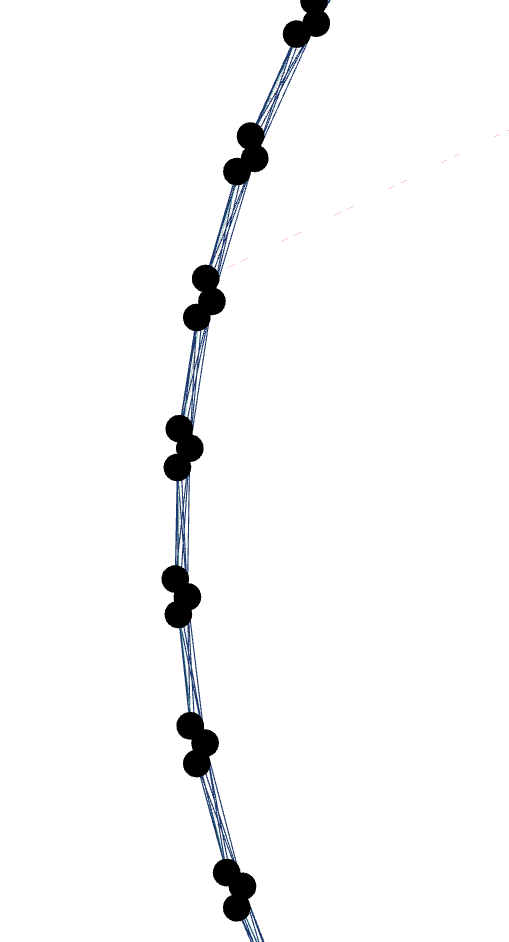}}
\subfloat[More Original Noise]{\includegraphics[width=0.716\linewidth]{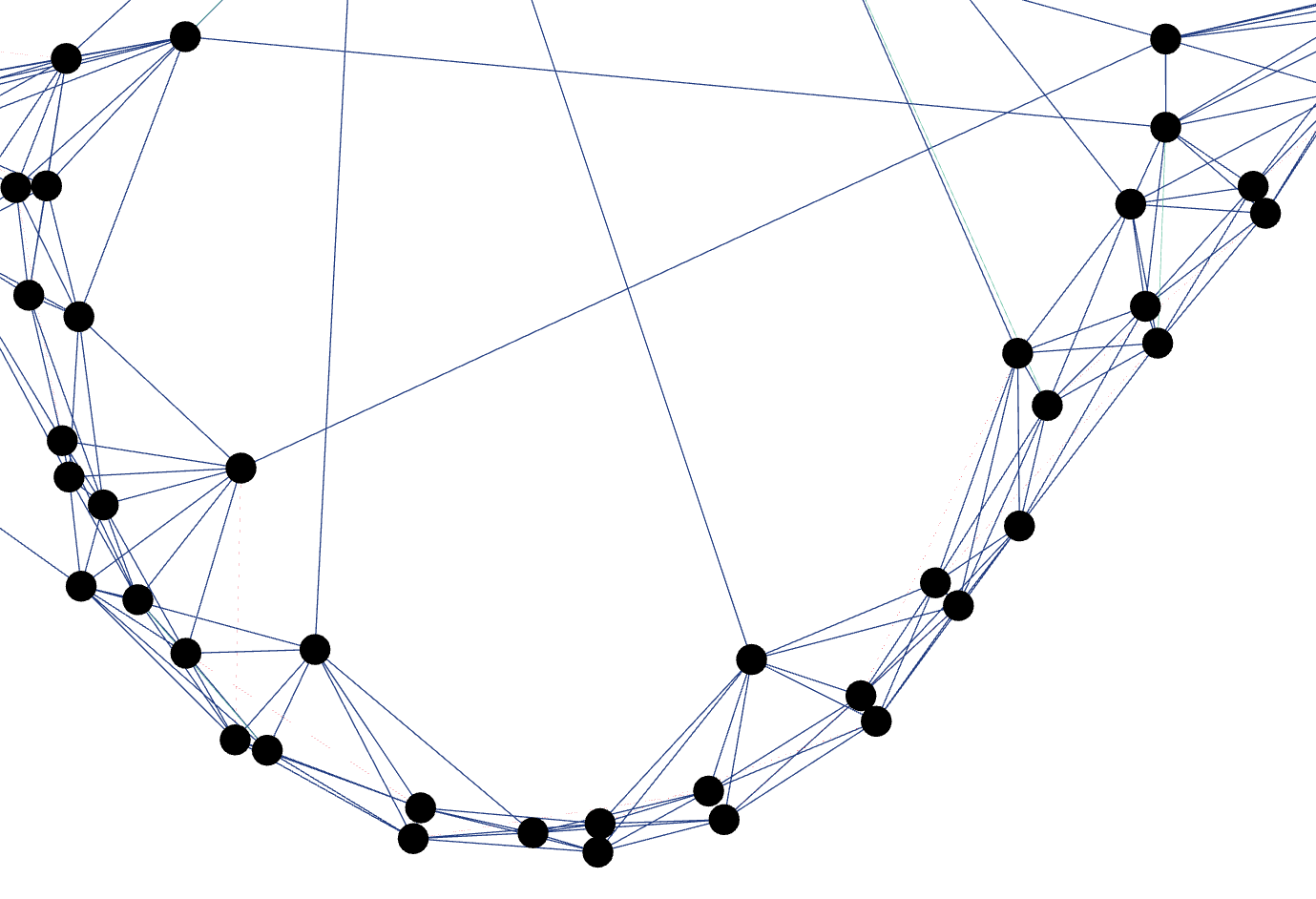}}
\caption{\textbf{Snapshots of Schema-Noise Decompositions for Randomly-Perturbed Wreath Graphs} -- Rather than restoring the graph to being a wreath, SCHENO GA instead points out that nodes in a wreath are nearly structurally identical to some of their neighbors; it modifies the structure accordingly to make them actually identical. \textbf{(a)} When less noise is present, the algorithm often makes triples of nodes equivalent to each other. \textbf{(b)} The criss-cross edges were due to the random noise. Here many pairs of nodes are made equivalent; they connect to exactly the same set of nodes.}\label{fig:wreath}
\end{figure}

\begin{figure*}[t]
\centering
\subfloat[What the Algorithm was Given]{\includegraphics[width=.48\textwidth]{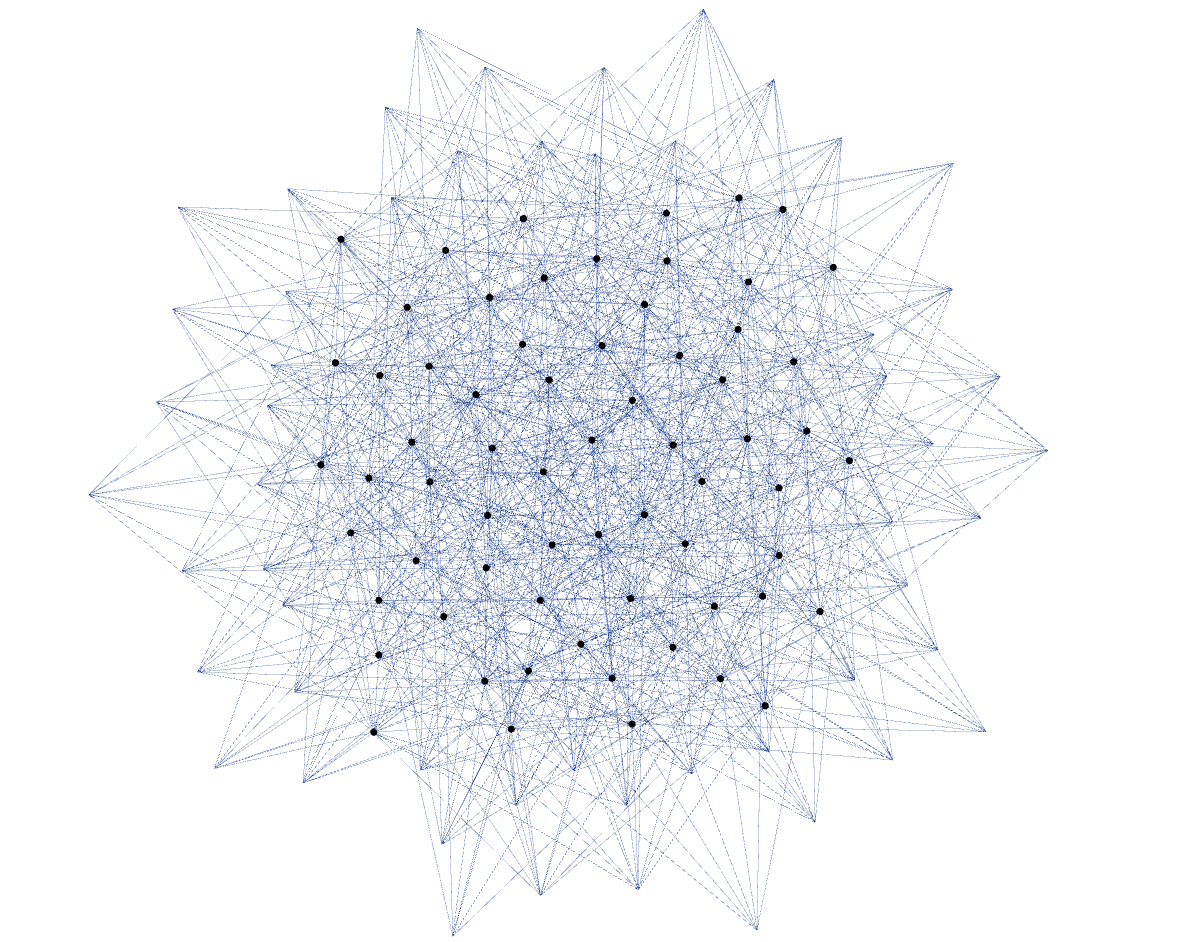}}
\subfloat[What it Found]{\includegraphics[width=.48\textwidth]{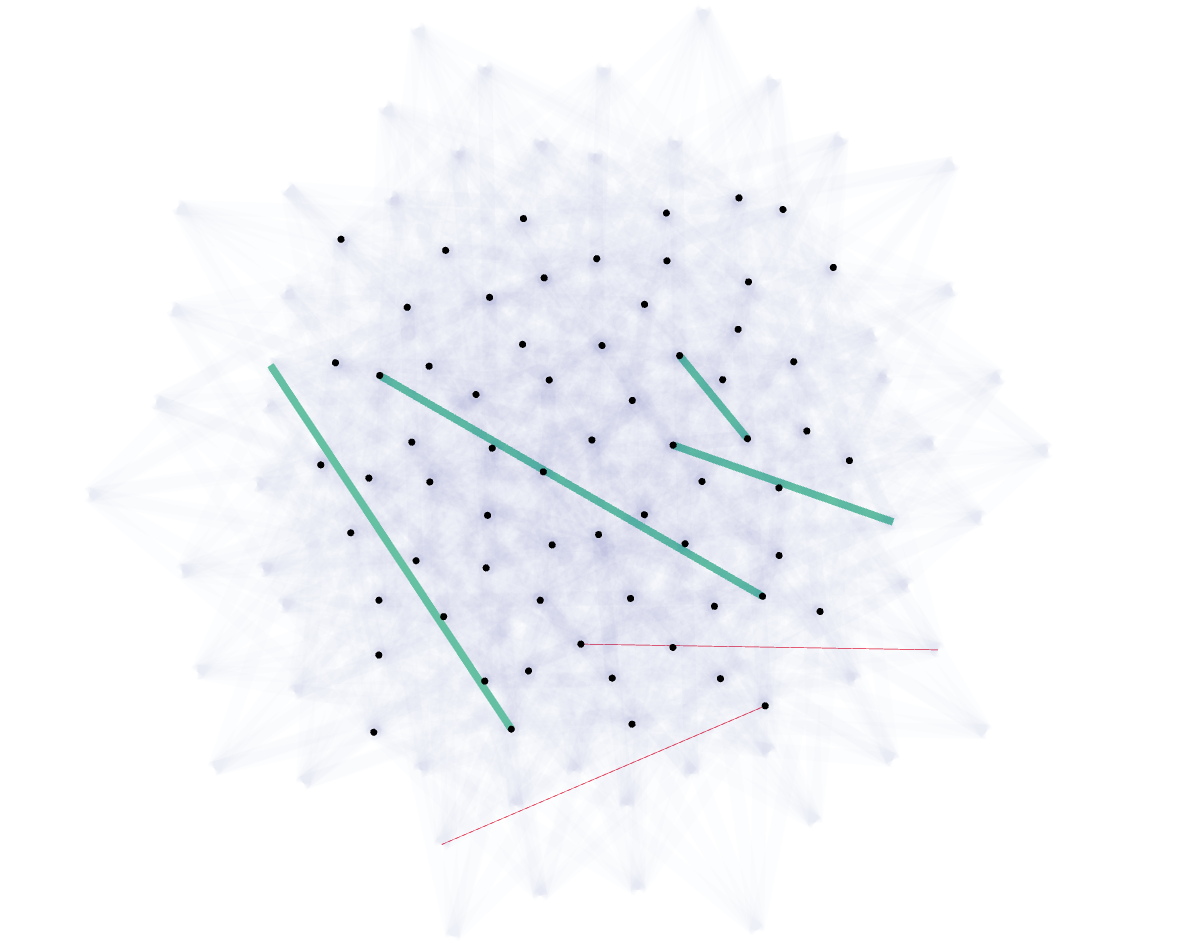}}
\caption{\textbf{Performance on the $(10, 3)$ Johnson Graph} -- On the left we see the highly structured form of a Johson graph, but with just 6 incorrect edges, it has only the identity automorphism. The right shows the missing edges as found by SCHENO GA. Teal means that the algorithm added the edges; red means the algorithm deleted them. Once these changes are made, the graph is restored to its 3,628,800 automorphisms.}\label{fig:johnson}
\end{figure*}

In the Johnson graph, the graph's symmetries are extremely interrelated to each other, when just a few (\eg\ 4 to 5) of the 1,260 edges are randomly perturbed, the graph loses almost \emph{all} of its symmetry (all of its original 3,628,800 automorphisms). When about 7 or fewer edges are removed, SCHENO GA recovers the structure, but when more are removed, the algorithm struggles. Crucially though, if the genetic algorithm would hypothetically manage to find the noise edges, SCHENO would give the true solution the highest score, but for a genetic algorithm with such a huge search space to explore, this is difficult to find without greedy gains along the way. To get a sense for what the algorithm needs to find we show one of SCHENO GA's successful runs in Figure~\ref{fig:johnson}.

\subsubsection{MNIST Image Data}

We now turn to one final kind of data: images. We look at the MNIST image dataset of black-and-white hand-written digits. An image is not clearly a graph, but we can convert a black-and-white image to a graph by treating the image as an adjacency matrix of a directed graph. As the MNIST digits are white on a black background, we do the conversion as follows:

If pixel $(i, j)$ is white, then we add directed edge $(i, j)$ to the graph when $j < i$, and we add directed edge $(i, j+1)$ to the graph when $j \geq i$. This use of $j+1$ avoids adding self-loops, which our code does not handle at present. To convert back from a graph to an image, we reverse this process; the presence of edge $(i, j)$ means pixel $(i, j)$ is white when $j < i$, and it means pixel $(i, j - 1)$ is white when $j > i$.

Overall, we were pleased with the results. We should stress that SCHENO GA is simply trying to find patterns and noise; it is \emph{not} trying to split handwritten digits into one of the 10 digit categories. As humans we've spent our lives learning to identify the relevant ``pattern'' as ``which of the 10 digits is this?'' but that is not the only way to see the data. In fact, that is a pattern \emph{across images of digits} more so than a pattern \emph{within} an image of a digit, though of course to recognize digits across images, we utilize sub-patterns like lines and curves within an image.

\begin{figure*}[t]
\centering
\mnistfig{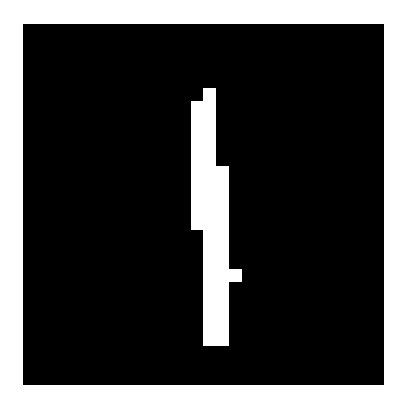}\hspace{1cm}
\mnistfig{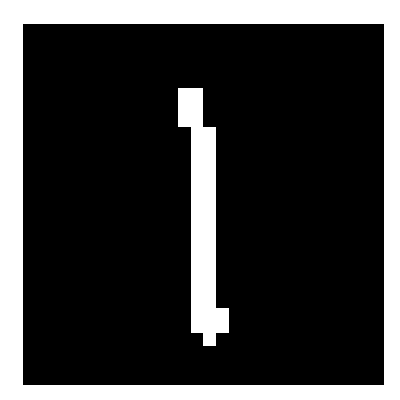}\\
\mnistfig{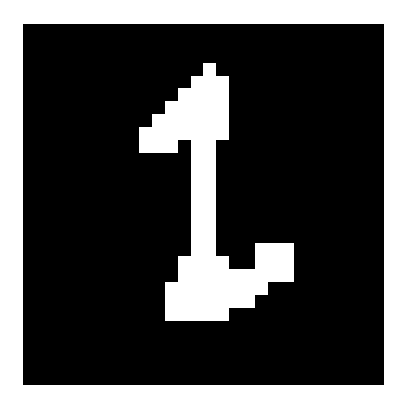}\hspace{1cm}
\mnistfig{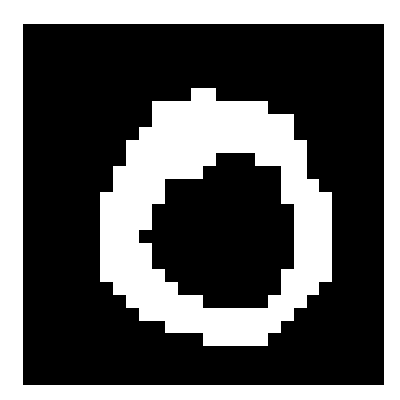}\\
\mnistfig{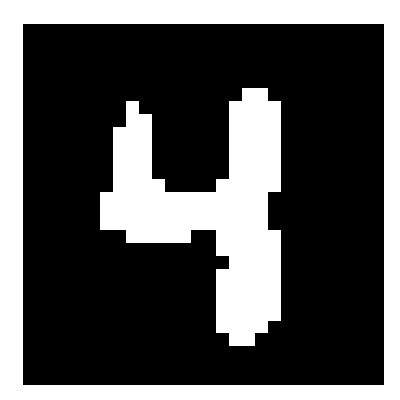}\hspace{1cm}
\mnistfig{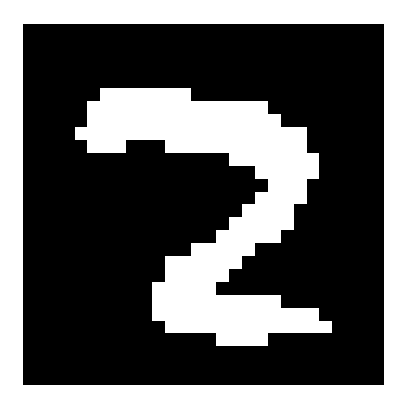}\\
\mnistfig{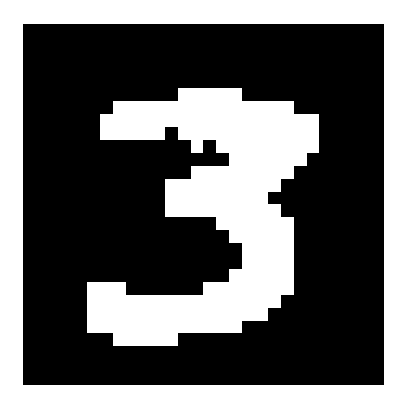}\hspace{1cm}
\mnistfig{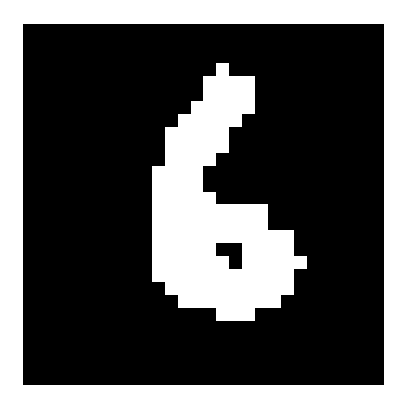}\\
\mnistfig{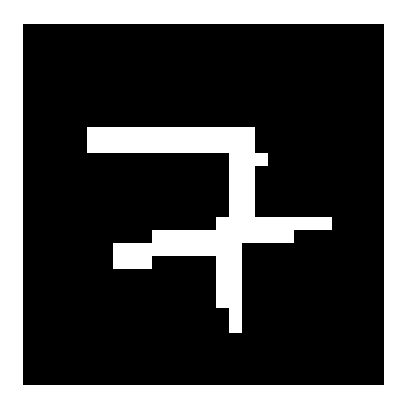}\hspace{1cm}
\mnistfig{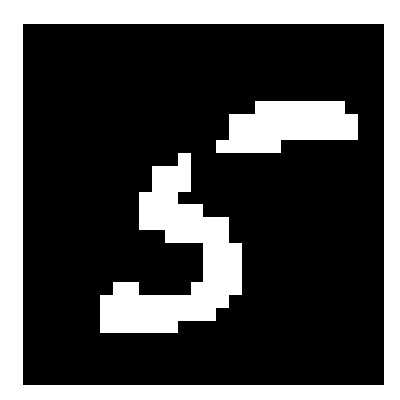}
\caption{\textbf{MNIST -- Our Favorite Results}}\label{fig:mnist_successes}
\end{figure*}

\begin{figure*}[t]
\centering
\mnistfig{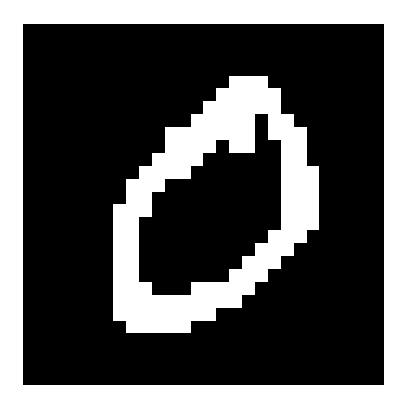}\hspace{1cm}
\mnistfig{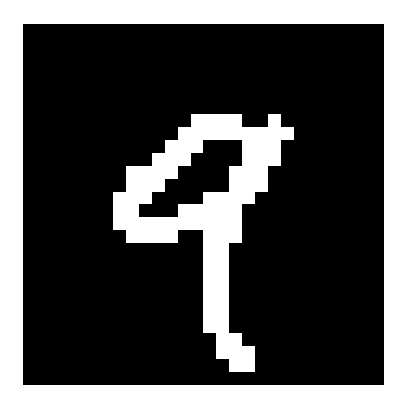}\\
\mnistfig{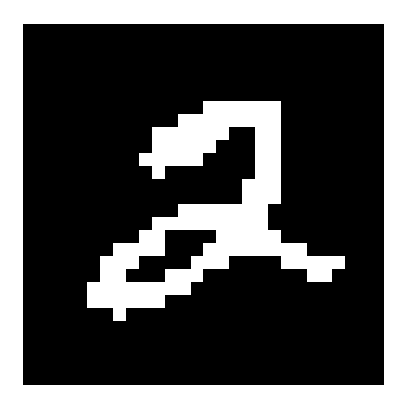}\hspace{1cm}
\mnistfig{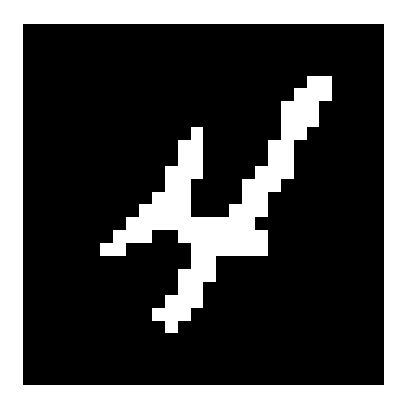}\\
\mnistfig{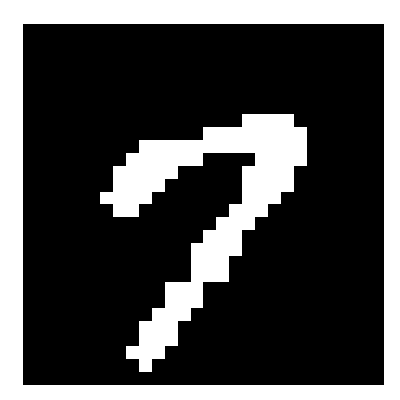}\hspace{1cm}
\mnistfig{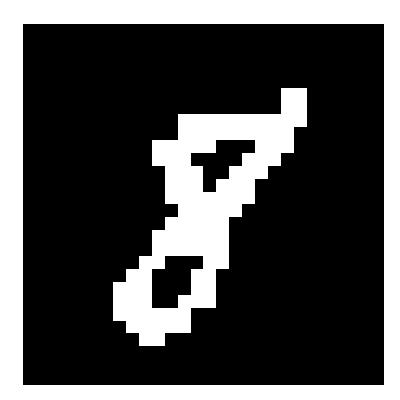}\\
\mnistfig{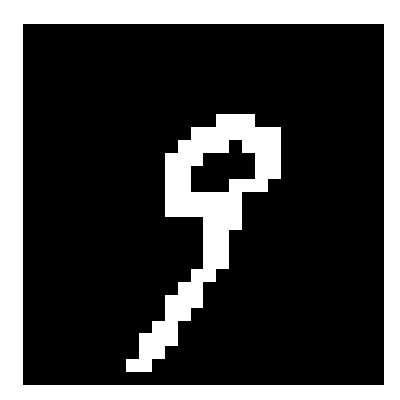}\hspace{1cm}
\mnistfig{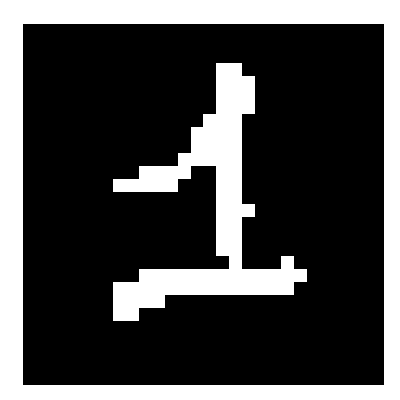}\\
\mnistfig{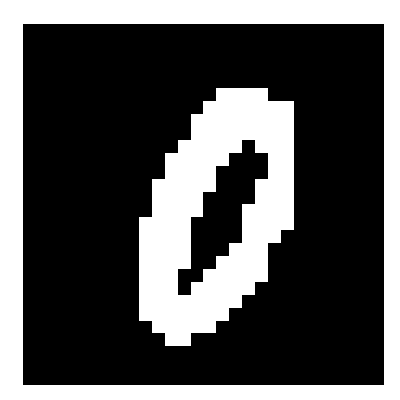}\hspace{1cm}
\mnistfig{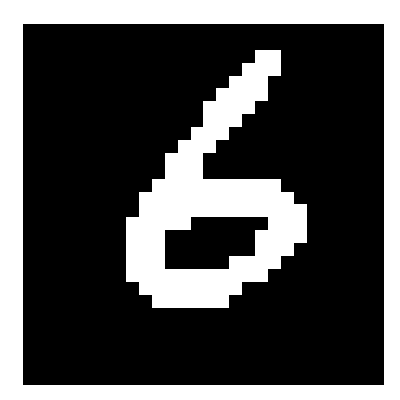}
\caption{\textbf{MNIST -- Typical Results}}\label{fig:mnist_normal}
\end{figure*}

\begin{figure*}[t]
\centering
\mnistfig{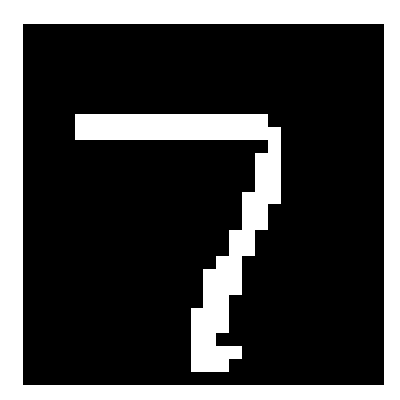}\hspace{1cm}
\mnistfig{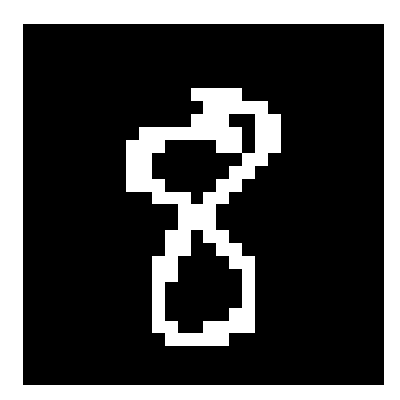}\\
\mnistfig{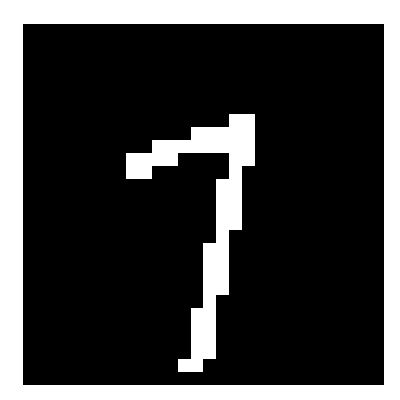}\hspace{1cm}
\mnistfig{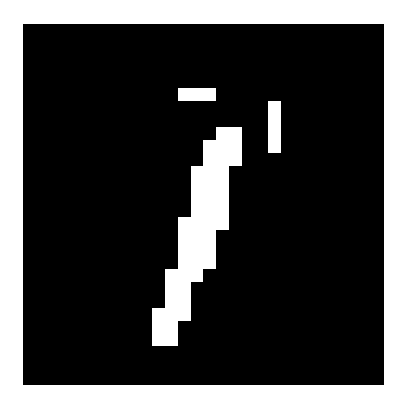}\\
\mnistfig{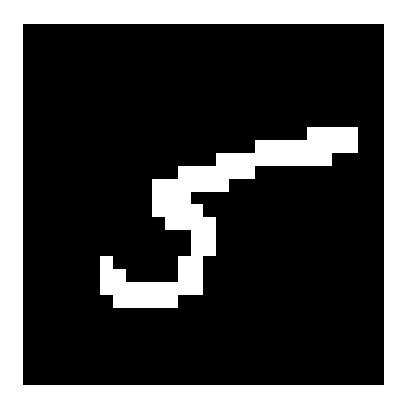}\hspace{1cm}
\mnistfig{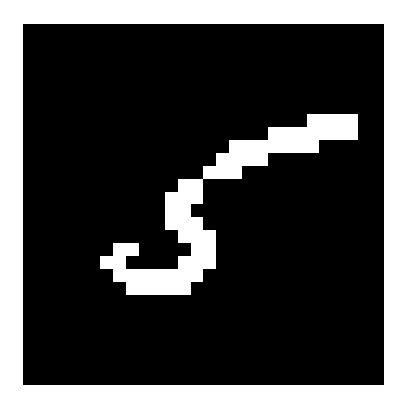}\\
\mnistfig{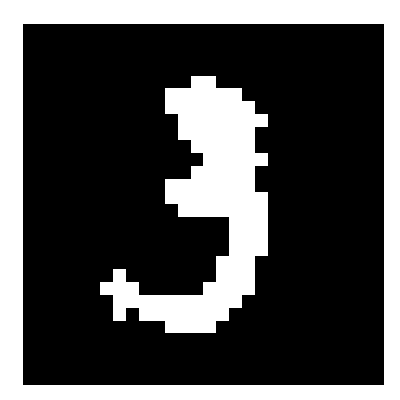}\hspace{1cm}
\mnistfig{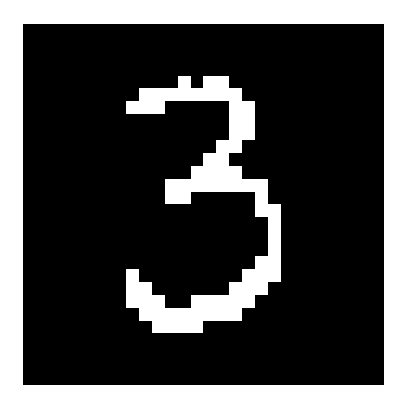}\\
\mnistfig{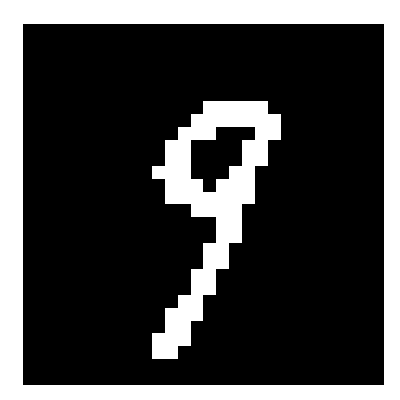}\hspace{1cm}
\mnistfig{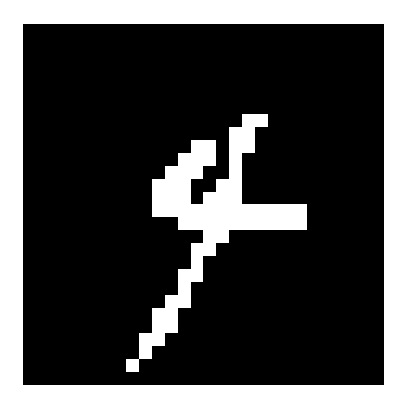}
\caption{\textbf{MNIST -- Strangest Results} -- The tendency to ``blockify'' continues. Some images with few white pixels are made even sparser. The changes we least understand are the last two shown.}\label{fig:mnist_weird}
\end{figure*}

In general, the main trend was to make images more block-like. Sometimes this makes the numeric digits more recognizeable qua digit to a human; sometimes this means finding a different kind of pattern/noise. The key takeaway is that our definition of schema and noise was extremely general and can unearth many kinds of patterns, even in visual data. Our computations were performed on 100 randomly-sampled MNIST digits. We show some representative results in Figures~\ref{fig:mnist_successes}, \ref{fig:mnist_normal}, and~\ref{fig:mnist_weird}.

\subsection{Training the GINs}\label{sec:GIN_parameters}

To train the graph isomorphism networks, we used the follow hyperparameters: Two layers, hidden dimension of 128, output size of 64. We used a learning rate of 0.001 decaying by a factor $\alpha = 0.9$ every 50 epochs, running for a total of 1000 epochs, with no dropout.

% To train the graph isomorphism networks, we used the follow hyperparameters: One layer with output dimension of 64. We used a learning rate of 0.01 decaying by a factor $\alpha = 0.5$ every 50 epochs, running for a total of 350 epochs, with no dropout.

We were analyzing networks without reference to any node features, so as our input ``features'' we give each of the $n$ nodes an $n$-length 1-hot vector: node $i$'s input is all zeros except index $i$ which is set to 1. This means that the input to the network is the identity matrix, which essentially allows the network to design its own node-level features. The network used the inner product decoder which means it produced a symmetric (\ie\ undirected) adjacency matrix.

Since our goal was not to evaluate the network in terms of link prediction as such but rather to get a structurally distilled version of the original graph, we did not set aside a set of validation or test edges; we used the full set of the graph's edges in training. Every epoch used a new randomly sampled set of negative edges equal in size to the set of true edges.

\end{document}